\documentclass{siamltex}

\usepackage{citesort}
\usepackage{epsfig}

\begin{document}

\newcommand{\gmu}{G-\{u\}}
\newcommand{\gm}[1]{G-\{#1\}}
\newcommand{\NC}{\Delta}
\newcommand{\kval}{{\rm val}} 
\newcommand{\mc}[2]{#1{\ominus}#2}  
\newcommand{\nR}{{R}} 
\newcommand{\lc}[2]{#1{\otimes}#2}
\newcommand{\mcc}[2]{{#1}\ominus{#2}}  
\newcommand{\Gz}{{{G}_z}}
\newcommand{\Gminus}[1]{{G}_z{-\{{#1}\}}}
\newcommand{\lcc}[2]{#1{\otimes}#2}
\newcommand{\HH}[0]{{\cal H}}
\newcommand{\AAA}[0]{{\cal A}}
\newcommand{\W}{}
\newcommand{\Wd}{}

\newtheorem{fact}[theorem]{Fact}
\newcommand{\mm}[0]{\mbox{\sc m}}
\newcommand{\mmb}[0]{\overline{\mbox{\sc m}}}

\newcommand{\kmast}{mast} 
\newcommand{\mast}{\mbox{\rm mast}}
\newcommand{\MAST}{\mbox{\rm mast}}
\newcommand{\Stree}{\mbox{\it S}}
\newcommand{\tree}{{\cal T}} 
\newcommand{\labels}[2]{{L}_{\scriptscriptstyle #1}(#2)}
\newcommand{\ralpha}[2]{{\alpha}_{\scriptscriptstyle #1}(#2)}
\newcommand{\cleaf}{\gamma}
\newcommand{\rbeta}[1]{\beta(#1)}
\newcommand{\rb}[1]{{\beta_{\scriptscriptstyle 1\succ 2}(#1)}}
\newcommand{\rbb}[1]{\beta_{\scriptscriptstyle 2\succ 1}(#1)}
\newcommand{\complement}[1]{\overline{#1}}

\newcommand{\mwm}{\mbox{\rm mwm}}
\newcommand{\rrmast}{\mbox{\rm r-mast}}

\newcommand{\funMNote}[1]{\varphi_{#1}} 

\newcommand{\children}{{\rm Ch}} 

\title{Cavity Matchings, Label Compressions, and Unrooted Evolutionary
Trees\thanks{A preliminary version appeared as part of {\em General
techniques for comparing unrooted evolutionary trees}, in Proceedings
of the~29th Annual ACM Symposium on Theory of Computing, 1997,
pp.~54--65, and part of {\em 
All-cavity Maximum Matchings}, in Proceedings of the~8th Annual
International Symposium on Algorithms and Computation, 1997, pp.~364-373.
}}
\author{Ming-Yang Kao\thanks{Department of Computer Science,
        Yale University, New Haven, CT 06520, U.S.A.,
        kao-ming-yang@cs.yale.edu.   Research
        supported in part by NSF Grant CCR-9531028.}
        \and
        Tak-Wah Lam\thanks{Department of Computer Science and
          Information Systems,
                The University of Hong Kong, Hong Kong,
                \{twlam, wksung, hfting\}@csis.hku.hk.  Research
                supported in part by Hong Kong RGC Grant 
                HKU-7027/98E.}
        \and
                Wing-Kin Sung\footnotemark[3]
        \and
                Hing-Fung~Ting\footnotemark[3]}
        \date{}

\maketitle

\begin{abstract}
We present an algorithm for computing a maximum agreement subtree of
two unrooted evolutionary trees.  It takes $O(n^{1.5} \log n)$ time
for trees with unbounded degrees, matching the best known time
complexity for the rooted case.  Our algorithm allows the input trees
to be mixed trees, i.e.,\ trees that may contain directed and
undirected edges at the same time.  Our algorithm adopts a recursive
strategy exploiting a technique called label compression.  The
backbone of this technique is an algorithm that computes the maximum
weight matchings over many subgraphs of a bipartite graph as fast as
it takes to compute a single matching.
\end{abstract}



\section{Introduction}
An {\it evolutionary} tree is one  whose leaves are labeled with distinct
symbols representing species.  Evolutionary trees are useful for modeling
the evolutionary relationship of species
\cite{AF94,ka94,BFW92,Gusfield91,KANNANLW1996,WJL96}.
An {\it agreement subtree} of two evolutionary trees
is an evolutionary tree that is also a topological subtree of the two
given trees.  A {\it maximum} agreement subtree is one with the largest
possible number of leaves.  Different models about the evolutionary
relationship of the same species may result in different evolutionary
trees.  A fundamental problem in computational biology is to determine how
much two models of evolution have in common.
To a certain extent, this problem can be
solved by computing a maximum agreement subtree of two given evolutionary
trees \cite{fg85}.

Algorithms for computing a maximum agreement subtree of two unrooted
evolutionary trees as well as two rooted trees have been studied
intensively in the past few years.  The unrooted case is more
difficult than the rooted case. There is indeed a linear-time
reduction from the rooted case to the unrooted one, but the reverse is
not known.
Steel and Warnow \cite{SW93}
gave the first polynomial-time algorithm for unrooted trees,
which runs in
$O(n^{4.5}\log{n})$ time.  Farach and Thorup reduced the time
to $O(n^{2+o(1)})$ for unrooted trees \cite{ft95.infcomp}
and $O(n^{1.5} \log n)$ for rooted
trees \cite{ftfocs}.
For the unrooted
case, the time  was improved by Lam, Sung and Ting \cite{lst} to
$O(n^{1.75 + o(1)})$.
Algorithms that work well for rooted trees with degrees bounded by a
constant have also been revealed recently.  The algorithm of Farach,
Przytycka and Thorup \cite{fpt95} takes $O(n \log^3 n)$ time,
and that of Kao
\cite{Kao.Evol.Siam} takes $O(n \log^2 n)$ time.
Cole and Hariharan \cite{ch96}  gave
an $O(n \log n)$-time algorithm
for
the case
where the input is further restricted to binary rooted trees.

This paper presents an algorithm for computing a maximum agreement
subtree of two unrooted trees.  It takes $O(n^{1.5} \log n)$ time for
trees with unbounded degrees, matching the best known time complexity
for the rooted case \cite{ftfocs}.  If the degrees are bounded by
a constant, the running time is only $O(n \log^4 n)$.  We omit the
details of this reduction since Przytycka \cite{Przytycka98} recently
devised an $O(n \log n)$-time algorithm for the same case.

Our algorithm allows the input trees to be {\em mixed\/} trees, i.e.,
trees that may contain directed and undirected edges at the same time
\cite{Gusfield87,Kao95}.  Such trees can handle a broader range of
information than rooted and unrooted trees.  To simplify the
discussion, this paper focuses on unrooted trees.
Our subtree algorithm
adopts a conceptually simple recursive strategy exploiting a novel
technique called {\em label compression}.  This technique enables our
algorithm to process overlapping subtrees iteratively while keeping
the total tree size very close to the original input size.  Label
compression builds on an unexpectedly fast algorithm for the {\em
all-cavity maximum weight matching\/} problem
\cite{klst-matching.scp}, which asks for the weight of a maximum
weight matching in $G-\{u\}$ for each node $u$ of a bipartite graph
$G$ with integer edge weights.  If $G$ has $n$ nodes, $m$ edges and
maximum edge weight $N$, the algorithm takes $O(\sqrt{n}m\log (n N))$
time, which matches the best known time bound for computing a single
maximum weight matching of $G$ due to Gabow and Tarjan \cite{GaTa}.

In \S\ref{all_cavity}, we solve the all-cavity matching problem.  In
\S\ref{sec_frame}, we formally define maximum agreement subtrees and
outline our recursive strategy for computing them.  We describe label
compression in \S\ref{sec_compress}, detail our subtree algorithm in
\S\ref{sec_mast_u1x_u2}, and discuss how to compute auxiliary
information for label compression in \S\ref{sec_aux_0} and
\S\ref{sec_aux_1_2}. We conclude by extending the subtree algorithm to
mixed trees in \S\ref{SectionMixedtree}.

\section{All-cavity maximum weight matching} \label{all_cavity}
Let $G=(X,Y,E)$ be a bipartite graph with $n$ nodes and $m$ edges
where each edge $(u, v)$ has a positive integer weight $w(u,v) \leq
N$.  Let $\mwm(G)$ denote the weight of a maximum weight matching in
$G$.  The all-cavity matching problem asks for $\mwm(\gmu)$ for all $u
\in X \cup Y$.  A naive
approach to solve this problem
is to compute $\mwm(\gm{u})$ separately for each $u$ using
the fastest algorithm for computing a single maximum weight matching
\cite{GaTa}, thus taking $O(n^{1.5} m \log (n N))$ total time.  A main
finding of this paper is that the matchings in different subgraphs
$\gmu$ are closely related and can be represented succinctly.  {From}
this representation, we can solve the problem in
$O(\sqrt{n}m\log(nN))$ time.  By symmetry, we only detail how to
compute $\mwm(\gm{u})$ for all $u \in X$.  Below we assume $m \geq
n/2$; otherwise, we remove the degree-zero nodes and work on the
smaller resultant graph.

A node $v$ of $G$ is {\em matched\/} by a matching of $G$ if $v$
is an endpoint of an edge in the matching.  In the remainder of this section,
let $M$ be a fixed maximum weight matching of $G$; also let $w(H)$ be
the total weight of a set $H$ of edges.  An {\em
alternating path\/} is a simple path ${P}$ in $G$ such that (1) $P$
starts with an edge in $M$, (2) the edges of $P$ alternate between $M$
and $E - M$, and (3) if $P$ ends at an edge $(u,v) \not\in M$, then $v$ is
not matched by $M$.  An {\em alternating cycle\/} is a simple cycle
$C$ in $G$ whose edges alternate between $M$ and $E - M$.  ${P}$
(respectively, $C$) can {\em transform\/} $M$ to another matching $M'
= {P} \cup M - {P} \cap M$ (respectively, $C \cup M - C \cap M$).  The
{\em net change\/} induced by ${P}$, denoted by $\NC(P)$, is
$w(M')-w(M)$, i.e., the total weight of the edges of ${P}$ in $E-M$
minus that of the edges of ${P}$ in $M$.  The {\em net change\/}
induced by $C$ is defined similarly.

\begin{figure}
\begin{center}
\begin{picture}(0,0)%
\epsfig{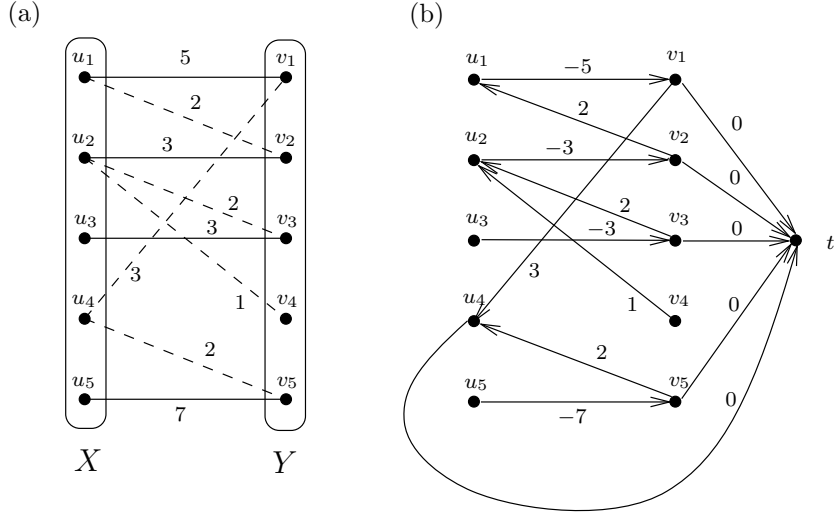}%
\end{picture}%
\setlength{\unitlength}{0.00083300in}%
\begingroup\makeatletter\ifx\SetFigFont\undefined
\def\x#1#2#3#4#5#6#7\relax{\def\x{#1#2#3#4#5#6}}%
\expandafter\x\fmtname xxxxxx\relax \def\y{splain}%
\ifx\x\y   
\gdef\SetFigFont#1#2#3{%
  \ifnum #1<17\tiny\else \ifnum #1<20\small\else
  \ifnum #1<24\normalsize\else \ifnum #1<29\large\else
  \ifnum #1<34\Large\else \ifnum #1<41\LARGE\else
     \huge\fi\fi\fi\fi\fi\fi
  \csname #3\endcsname}%
\else
\gdef\SetFigFont#1#2#3{\begingroup
  \count@#1\relax \ifnum 25<\count@\count@25\fi
  \def\x{\endgroup\@setsize\SetFigFont{#2pt}}%
  \expandafter\x
    \csname \romannumeral\the\count@ pt\expandafter\endcsname
    \csname @\romannumeral\the\count@ pt\endcsname
  \csname #3\endcsname}%
\fi
\fi\endgroup
\begin{picture}(5206,3182)(441,-2693)
\put(3044,375){\makebox(0,0)[b]{\smash{\SetFigFont{10}{12.0}{rm}(b)}}}
\put(4280,-815){\makebox(0,0)[b]{\smash{\SetFigFont{8}{9.6}{rm}$2$}}}
\put(4326,-1429){\makebox(0,0)[b]{\smash{\SetFigFont{8}{9.6}{rm}$1$}}}
\put(4135,-1727){\makebox(0,0)[b]{\smash{\SetFigFont{8}{9.6}{rm}$2$}}}
\put(3355,-1909){\makebox(0,0)[b]{\smash{\SetFigFont{8}{9.6}{rm}$u_5$}}}
\put(3343,-1378){\makebox(0,0)[b]{\smash{\SetFigFont{8}{9.6}{rm}$u_4$}}}
\put(3362,-884){\makebox(0,0)[b]{\smash{\SetFigFont{8}{9.6}{rm}$u_3$}}}
\put(3356,-372){\makebox(0,0)[b]{\smash{\SetFigFont{8}{9.6}{rm}$u_2$}}}
\put(3362,116){\makebox(0,0)[b]{\smash{\SetFigFont{8}{9.6}{rm}$u_1$}}}
\put(4608,140){\makebox(0,0)[b]{\smash{\SetFigFont{8}{9.6}{rm}$v_1$}}}
\put(4621,-365){\makebox(0,0)[b]{\smash{\SetFigFont{8}{9.6}{rm}$v_2$}}}
\put(4621,-878){\makebox(0,0)[b]{\smash{\SetFigFont{8}{9.6}{rm}$v_3$}}}
\put(4620,-1384){\makebox(0,0)[b]{\smash{\SetFigFont{8}{9.6}{rm}$v_4$}}}
\put(4621,-1897){\makebox(0,0)[b]{\smash{\SetFigFont{8}{9.6}{rm}$v_5$}}}
\put(4134,-960){\makebox(0,0)[b]{\smash{\SetFigFont{8}{9.6}{rm}$-3$}}}
\put(3952,-2132){\makebox(0,0)[b]{\smash{\SetFigFont{8}{9.6}{rm}$-7$}}}
\put(3870,-449){\makebox(0,0)[b]{\smash{\SetFigFont{8}{9.6}{rm}$-3$}}}
\put(4980,-960){\makebox(0,0)[b]{\smash{\SetFigFont{8}{9.6}{rm}$0$}}}
\put(3715,-1219){\makebox(0,0)[b]{\smash{\SetFigFont{8}{9.6}{rm}$3$}}}
\put(3984, 53){\makebox(0,0)[b]{\smash{\SetFigFont{8}{9.6}{rm}$-5$}}}
\put(4022,-205){\makebox(0,0)[b]{\smash{\SetFigFont{8}{9.6}{rm}$2$}}}
\put(5563,-1042){\makebox(0,0)[b]{\smash{\SetFigFont{8}{9.6}{rm}$t$}}}
\put(4942,-2022){\makebox(0,0)[b]{\smash{\SetFigFont{8}{9.6}{rm}$0$}}}
\put(4962,-1434){\makebox(0,0)[b]{\smash{\SetFigFont{8}{9.6}{rm}$0$}}}
\put(4968,-637){\makebox(0,0)[b]{\smash{\SetFigFont{8}{9.6}{rm}$0$}}}
\put(4980,-301){\makebox(0,0)[b]{\smash{\SetFigFont{8}{9.6}{rm}$0$}}}
\put(1691,-1713){\makebox(0,0)[b]{\smash{\SetFigFont{8}{9.6}{rm}$2$}}}
\put(1881,-1415){\makebox(0,0)[b]{\smash{\SetFigFont{8}{9.6}{rm}$1$}}}
\put(1508,-2118){\makebox(0,0)[b]{\smash{\SetFigFont{8}{9.6}{rm}$7$}}}
\put(1602,-169){\makebox(0,0)[b]{\smash{\SetFigFont{8}{9.6}{rm}$2$}}}
\put(1425,-435){\makebox(0,0)[b]{\smash{\SetFigFont{8}{9.6}{rm}$3$}}}
\put(1539,109){\makebox(0,0)[b]{\smash{\SetFigFont{8}{9.6}{rm}$5$}}}
\put(1836,-801){\makebox(0,0)[b]{\smash{\SetFigFont{8}{9.6}{rm}$2$}}}
\put(906,-1897){\makebox(0,0)[b]{\smash{\SetFigFont{8}{9.6}{rm}$u_5$}}}
\put(906,-1390){\makebox(0,0)[b]{\smash{\SetFigFont{8}{9.6}{rm}$u_4$}}}
\put(919,-885){\makebox(0,0)[b]{\smash{\SetFigFont{8}{9.6}{rm}$u_3$}}}
\put(906,-385){\makebox(0,0)[b]{\smash{\SetFigFont{8}{9.6}{rm}$u_2$}}}
\put(906,123){\makebox(0,0)[b]{\smash{\SetFigFont{8}{9.6}{rm}$u_1$}}}
\put(2171,115){\makebox(0,0)[b]{\smash{\SetFigFont{8}{9.6}{rm}$v_1$}}}
\put(2165,-384){\makebox(0,0)[b]{\smash{\SetFigFont{8}{9.6}{rm}$v_2$}}}
\put(2177,-885){\makebox(0,0)[b]{\smash{\SetFigFont{8}{9.6}{rm}$v_3$}}}
\put(2177,-1389){\makebox(0,0)[b]{\smash{\SetFigFont{8}{9.6}{rm}$v_4$}}}
\put(2177,-1896){\makebox(0,0)[b]{\smash{\SetFigFont{8}{9.6}{rm}$v_5$}}}
\put(1229,-1245){\makebox(0,0)[b]{\smash{\SetFigFont{8}{9.6}{rm}$3$}}}
\put(1704,-947){\makebox(0,0)[b]{\smash{\SetFigFont{8}{9.6}{rm}$3$}}}
\put(932,-2427){\makebox(0,0)[b]{\smash{\SetFigFont{12}{14.4}{rm}$X$}}}
\put(2159,-2440){\makebox(0,0)[b]{\smash{\SetFigFont{12}{14.4}{rm}$Y$}}}
\put(534,381){\makebox(0,0)[b]{\smash{\SetFigFont{10}{12.0}{rm}(a)}}}
\end{picture}
 \end{center}
 \caption{$($a$)$ a bipartite graph $G$;
 $($b$)$ the corresponding directed graph $D$.}
   \label{transform-fig}
 \end{figure}

The next lemma divides the computation of $\mwm(\gm{u})$ into two
cases.
\begin{lemma}\label{lem_match_nonmatch} Let $u \in X$.
\begin{enumerate}
\item \label{prop1} If $u$ is not matched by $M$, then $M$ is
also a maximum weight matching in $\gm{u}$ and $\mwm(\gm{u})=\mwm(G)$.
\item\label{alternating-path-lemma} If $u$ is matched by $M$, then $G$
contains an alternating path $P$ starting from $u$, which can
transform $M$ to a maximum weight matching in $\gmu$.
\end{enumerate}
\end{lemma}
\begin{proof}
Statement 1 is straightforward.  To prove Statement 2, let $M'$ be a
maximum weight matching in $\gmu$.  Consider the
edges in $M \cup M' - M \cap M'$.  They form a set $S$ of alternating
paths and cycles.  Since $u$ is
matched by $M$ but not by $M'$, $u$ is of degree one in $M \cup M' - M
\cap M'$.  Let $P$ be the alternating path in $S$ with $u$ as an
endpoint.  Let $M''$ be the matching obtained by transforming $M$ only
with $P$.  Since $u$ is not matched by $M''$, $M''$ is a matching in
$\gmu$.  $M'$ can be obtained by further transforming $M''$ with the
remaining alternating paths and cycles in $S$.  The net change induced
by each of these alternating paths and cycles is non-positive;
otherwise, such a path or cycle can improve $M$ and we obtain a
contradiction.  Therefore, $w(M'') \geq w(M')$, i.e., both $M'$ and
$M''$ are maximum weight matchings in $\gmu$.
\end{proof}

By Lemma~\ref{lem_match_nonmatch}(\ref{alternating-path-lemma}), we
can compute $\mwm(\gmu)$ for any $u \in X$ matched by $M$ by finding
the alternating path starting from $u$ with the largest net change.  
Below we construct a
directed graph $D$, which enables us to identify such an
alternating path for every node easily.  The node set of $D$ is $X
\cup Y \cup \{t\}$, where $t$ is a new node.  The edge set of $D$ is
defined as follows; see Figure~\ref{transform-fig} for an example.
\begin{itemize}
\item If $x \in X$ is not matched by $M$,
$D$ has an edge from $x$ to $t$ with weight zero.
\item If $y \in Y$ is matched by $M$,
$D$ has an edge from $y$ to $t$ with weight zero.
\item If $M$ has an edge $(x, y)$ where $x \in X$ and $y \in Y$,
$D$ has an edge from $x$ to $y$ with weight $-w(x, y)$.
\item  If $E - M$ has an edge $(x, y)$ where $x \in X$ and $y \in Y$,
$D$ has an edge from $y$ to $x$ with weight $w(x, y)$.
\end{itemize}

Note that $D$ has $n+1$ nodes and at most $n+m$ edges.  The weight of
each edge in $D$ is an integer in $[-N, N]$.

\begin{lemma}\label{D-relate-G}
\begin{enumerate}
\item $D$ contains no positive-weight cycle.
\item \label{D-relate-G-2}
Each alternating path $P$ in $G$ that starts
from $u \in X$ corresponds to a simple path $Q$ in $D$ from $u$ to
$t$, and vice versa.  Also, $\NC(P) = w(Q)$.
\item \label{corollary1} For each $u \in X$ matched by $M$,
$\mwm(\gmu)$ is the sum of $\mwm(G)$ and the weight of the longest
path in $D$ from $u$ to $t$.
\end{enumerate}
\end{lemma}
\begin{proof}

Statement 1. Consider a simple cycle $C = u_1, u_2, \cdots, u_k, u_1$
in $D$.  Since $t$ has no outgoing edges, no $u_i$ equals $t$.  By the
definition of $D$, $C$ is also an alternating cycle in $G$.
Therefore, $w(C)$ is the net change induced by transforming $M$ with
$C$.  Since $M$ is a maximum weight matching in $G$, this net change
is non-positive.

Statement 2.  Consider an alternating path $P = u,u_1,u_2,\cdots,u_k$
in $G$ starting from $u$.  In $D$, $P$ is also a simple path.  If $u_k
\in X$, then $u_k$ is not matched by $M$, and $D$ contains the edge
$(u_k,t)$.  If $u_k \in Y$, then $u_k$ is matched by $M$, and $D$
again contains the edge $(u_k,t)$.  Therefore, $D$ contains the simple
path $Q = u, u_1, u_2, \cdots, u_k, t$.  The weight of $Q$ is
$\NC(P)$.  The reverse direction of the statement is straightforward.

Statement 3. This statement follows from
Lemma~\ref{lem_match_nonmatch}(\ref{alternating-path-lemma}) and
Statement \ref{D-relate-G-2}.
\end{proof}

\begin{theorem} \label{1stTheorem}
Given $G$, we can compute $\mwm(\gmu)$ for all nodes $u \in G$ in
$O(\sqrt{n} m \log(nN))$ time.
\end{theorem}
\begin{proof}
By symmetry and Lemmas~\ref{lem_match_nonmatch}(\ref{prop1}) and
\ref{D-relate-G}(\ref{corollary1}), we compute $\mwm(\gmu)$ for all $u
\in X$ as follows.
\begin{enumerate}
\item  Compute a maximum weight matching $M$ of $G$.
\item Construct $D$ as above and find the weights of its longest paths
to $t$.
\item For each $u \in X$, if $u$ is matched by $M$, then $\mwm(\gmu)$
is the sum of $\mwm(G)$ and the weight of the longest path from $u$ to
$t$ in $D$; otherwise, $\mwm(\gmu)=\mwm(G)$.
\end{enumerate}
Step 1 takes $O(\sqrt{n} m \log(nN))$ time.  At Step 2, constructing
$D$ takes $O(n+m)$ time, and the single-destination longest paths
problem takes $O(\sqrt{n} m \log N)$ time \cite{Goldberg95}.  Step 3
takes $O(n)$ time.  Thus, the total time is $O(\sqrt{n}m\log(nN))$.
\end{proof}

\section{The main result}\label{sec_frame}
This section gives a formal definition of maximum agreement subtrees
and an overview of our new subtree algorithm.

\subsection{Basics}
Throughout the remainder of this paper,
unrooted trees are denoted by $U$ or $X$, and rooted trees by
$T$, $W$ or $R$.
A node of degree 0 or 1 is a \emph{leaf}; otherwise, it is {\it
internal}.  Adopted to avoid technical trivialities, this definition
is somewhat nonstandard in that if the root of a rooted tree is of
degree 1, it is also a leaf.

For an unrooted tree $U$ and a node $u \in U$,
let $U^u$ denote the rooted tree constructed by rooting
$U$ at $u$.
For a rooted tree $T$ and a node $v \in T$, let $T^v$ denote the
{\em rooted subtree} of $T$ that comprises $v$ and its descendants.
Similarly, for a node $v \in U^{u}$, $U^{uv}$ is the
rooted subtree of $U^u$ rooted at $v$, which is also called a {\it rooted
subtree} of $U$.

An {\it evolutionary} tree
is a tree whose leaves are labeled with
distinct symbols.
Let $T$ be a rooted evolutionary tree with leaves labeled over a set $L$.
A label subset $L'\subseteq L$ {\it induces} a subtree of $T$, denoted by $T |
L'$, whose nodes are the leaves of $T$ labeled over $L'$
as well as the least
common ancestors of such leaves in $T$, and whose edges preserve the
ancestor-descendant relationship of $T$.  Consider two rooted evolutionary
trees $T_1$ and $T_2$ labeled over $L$.  Let $T'_1$ be a subtree of $T_1$
induced by some subset of $L$. We similarly define $T'_2$ for $T_2$.  If
there exists an isomorphism between $T'_1$ and $T'_2$ mapping each leaf in
$T'_1$ to one in $T'_2$ with the same label, then $T'_1$ and $T'_2$ are
each called {\em agreement subtrees} of $T_1$ and $T_2$.  
Note that this isomorphism is unique.  Consider any nodes $u \in T_1$
and $v \in T_2$.  We say that {\it $u$ is mapped to $v$ in} $T'_1$ and $T'_2$
if this isomorphism maps $u$ to $v$.  A {\em maximum} agreement
subtree of $T_1$ and $T_2$ is one containing the largest possible number of
labels.  Let $\mast(T_1, T_2)$ denote the number of labels in such a tree.
A {\it maximum agreement subtree} of two unrooted evolutionary trees $U_1$ and
$U_2$ is one with the largest number of labels among the maximum agreement
subtrees of $U_1^u$ and $U_2^v$ over all nodes $u \in U_1$ and $v \in U_2$.
Let
\begin{eqnarray}\label{eqn_mast}
  \mast(U_1, U_2) &=& \max\{\mast(U_1^u, U_2^v) \mid u \in U_1, v \in
  U_2\}.
\end{eqnarray}
{\it Remark.} The nodes $u$ (or $v$) can be restricted to internal nodes
when the trees have at least three nodes. We can also generalize the above
definition to handle a pair of rooted tree and unrooted tree $(T,U)$. That
is, $\mast(T, U)$ is defined to be $\max\{\mast(T, U^v) \mid v \in U\}$.

\subsection{Our subtree algorithm}
The next theorem is our main result.  
The {\it size} $|U|$ (or $|T|$)
of an unrooted tree $U$ (or a rooted tree $T$) is its node count.
\begin{theorem}\label{overall-analysis}
Let $U_1$ and $U_2$ be two unrooted evolutionary trees.
We can compute $\MAST(U_1, U_2)$ in $O(N^{1.5} \log N)$
time, where $N = \max \{|U_1|, |U_2|\}$.
\end{theorem}

We prove Theorem~\ref{overall-analysis} by presenting our algorithm in
a top-down manner with an outline here.
As in previous work, our algorithm only computes $\mast(U_1,U_2)$ 
and can be augmented to report a corresponding
subtree.  It uses graph separators.  A {\em separator} of a tree is an
internal node whose removal divides the tree into connected components
each containing at most half of the tree's nodes.  Every tree that
contains at least three nodes has a separator, which can be found in
linear time.

If $U_1$ or $U_2$ has at most two nodes, $\mast(U_1,U_2)$ as defined in
Equation~(\ref{eqn_mast}) can easily be computed in $O(N)$ time.
Otherwise, both
trees have at least three nodes each, and we can find a separator $x$ of
$U_1$.  We then consider three cases.

{\it Case 1}: In some maximum agreement subtree of $U_1$ and $U_2$, the
node $x$ is mapped to a node $y \in U_2$.  In this case,
$\MAST(U_1,U_2)=\mast(U_1^x, U_2)$.
To compute $\mast(U_1^x, U_2)$, we
might simply evaluate $\mast(U_1^x, U_2^y)$ for different $y$ in $U_2$.
This approach involves solving the $\mast$ problem for $\Theta(N)$
different pairs of rooted trees and introduces much redundant computation.
For example, consider a rooted subtree $R$ of $U_2$.  For all $y \in
U_2-R$, $R$ is a common subtree of $U_2^y$.  Hence, $R$ is examined
repeatedly in the computation of $\mast(U_1^x, U_2^y)$ for these $y$.  To
speed up the computation, we devise the technique of label compression in
\S\ref{sec_compress} to elicit sufficient information between $U_1^x$ and
$R$ so that we can compute $\mast(U_1^x, U_2^y)$ for all $y \in U_2 - R$
without examining $R$. 
This leads to an efficient algorithm for handling Case 1, the
time complexity is stated in the following lemma.

\begin{lemma} \label{mainlemma}
  Assume that $U_1$ and $U_2$ have at least three nodes each.  Given an
  internal node $x \in U_1$, we can compute $\mast(U_1^x, U_2)$ in
  $O(N^{1.5} \log N)$
  time.
\end{lemma}
\begin{proof}
See \S\ref{sec_compress} to \S\ref{sec_aux_1_2}. 
\end{proof}

{\it Case 2}: In some maximum agreement subtree of $U_1$ and $U_2$, two
certain nodes $v_1$ and $v_2$ of $U_1$ are mapped to nodes in $U_2$, and
$x$ is on the path in $U_1$ between $v_1$ and $v_2$.  This case
is similar to Case 1.  Let $\tilde{U}_2$ be the tree
constructed by adding a dummy node  in the middle of every edge
in $U_2$.  Then, $\MAST(U_1, U_2)=\mast(U_1^x,\tilde{U}_2^y)$ for some
dummy node $y$ in $\tilde{U}_2$.
Thus, $\MAST(U_1, U_2) = \mast(U_1^x,\tilde{U}_2)$.
As in Case 1, $\mast(U_1^x, \tilde{U}_2)$
can be computed in
$O(N^{1.5} \log N)$
time.

{\it Case 3}: None of the above two cases.  Let
$U_{1,1},U_{1,2},\ldots,U_{1,b}$ be the evolutionary trees formed by the
connected components of $U_1-\{x\}$.  Let $J_1,\ldots,J_b$ be the sets
of labels in these components, respectively.
Then, a maximum agreement subtree of $U_1$ and
$U_2$ is labeled over some $J_i$.  Therefore, $\MAST(U_1, U_2) = \max\{
\MAST(U_{1,i}, U_2 | J_i) \mid i \in [1, b]\}$, and we compute each
$\MAST(U_{1,i}, U_2 | J_i)$ recursively.

\newlength{\mini}
\setlength{\mini}{\textwidth}
\addtolength{\mini}{-11pt}
\begin{figure}
\begin{center}
\fbox{%
\begin{minipage}{\mini}
\begin{list}{}{\setlength{\leftmargin}{0.5em}}
 \item /* $U_1$ and $U_2$ are unrooted trees. */
 \item {$\mast(U_1, U_2)$}
 \begin{list}{}{\setlength{\leftmargin}{1em}}
  \item find a separator $x$ of $U_1$;
  \item construct $\tilde{U}_2$ by adding a dummy node $w$ at the middle of
      each edge $(u, v)$ in $U_2$;
  \item $\kval = \mast(U_1^x, U_2)$;
  \item $\kval' = \mast(U_1^x, \tilde{U}_2)$;
  \item let $U_{1,1}, U_{1,2}, \ldots, U_{1,b}$ be
       the connected components of $U_1 - \{x\}$;
  \item for all $i \in [1,b]$, let $J_i$ be the set of labels of $U_{1,i}$;
  \item {for all} $i \in [1,b]$, set $\kval_i = \mast(U_{1,i}, U_2|J_i) \}$;
  \item return $\max \{ \kval, \kval', \max_{1 \leq i \leq b}\kval_i \}$;
 \end{list}
\end{list}
\end{minipage}
} 
\end{center}
\caption{Algorithm for computing $\mast(U_1, U_2)$.} \label{algo1}
\end{figure}
Figure~\ref{algo1} summarizes the steps for computing $\mast(U_1,U_2)$.
Here we analyze the time complexity $T(N)$ based on Lemma~\ref{mainlemma}.
Cases 1 and 2 each take
$O(N^{1.5} \log N)$
time.  Let $N_i = |U_{1,i}|$.
Then Case 3 takes $\sum_{i \in [1,b]} T(N_i)$ time.  By recursion,
\[T(N) = O(N^{1.5} \log N) + \sum_{i \in [1,b]} T(N_i).\]
Since $x$ is a separator of $U_1$, $N_i \leq \frac{N}{2}$.  Then,
since $\sum_{i \in [1,b]} N_i \leq N$, $T(N) = O(N^{1.5} \log N)$
\cite{bhs80,kao.recur.ics}
and
the time bound in Theorem~\ref{overall-analysis} follows.  To complete
the proof of Theorem~\ref{overall-analysis}, we devote
\S\ref{sec_compress} through \S\ref{sec_aux_1_2} to proving
Lemma~\ref{mainlemma}.

\section{Label compressions} \label{sec_compress}
To compute a maximum agreement subtree, our algorithm recursively processes
overlapping subtrees of the input trees.  The technique of label
compression compresses overlapping parts of such subtrees to reduce their
total size.  We define label compressions with respect to a rooted subtree
in \S\ref{subsec_compress_one} and with respect to two label-disjoint
rooted subtrees in \S\ref{subsec_compress_two}.  We do not use label
compression with respect to three or more trees.

As a warm-up, let us define a concept called {\em subtree shrinking},
which is a primitive form of label compression.
Let $T$ be a rooted tree.  Let $R$ be a rooted subtree of $T$.  Let
$\mc{T}{R}$ denote the rooted tree obtained by replacing $R$ with a leaf
$\cleaf$.  We say that $\cleaf$ is a {\em shrunk} leaf. 
The other leaves are {\em atomic} leaves.  Similarly, for two
label-disjoint rooted subtrees $R_1$ and $R_2$ of $T$, let $\mc{T}{(R_1,
  R_2)}$ denote the rooted tree obtained by replacing $R_1$ and $R_2$ with
{\it shrunk} leaves $\cleaf_1$ and $\cleaf_2$, respectively.  We extend
these notions to an unrooted tree $U$ and define $\mc{U}{R}$ and
$\mc{U}{(R_1,R_2)}$ similarly.

\subsection{Label compression with respect to one rooted subtree}
\label{subsec_compress_one}
Let $T$ be a rooted tree.
Let $v$ be a node in $T$ and $u$ an ancestor of $v$.  Let $P$ be the path
of $T$ from $u$ to $v$.  A node {\it lies} between $u$ and $v$ if it is in
$P$ but differs from $u$ and $v$.  A subtree of $T$ is {\it attached} to
$u$ if it is some $T^w$ where $w$ is a child of $u$.  A subtree of $T$ {\it
  hangs} between $u$ and $v$ if it is attached to some node lying between
$u$ and $v$, but its root is not in $P$ and is not $v$.

\begin{figure}
\begin{center}
\begin{picture}(0,0)%
\epsfig{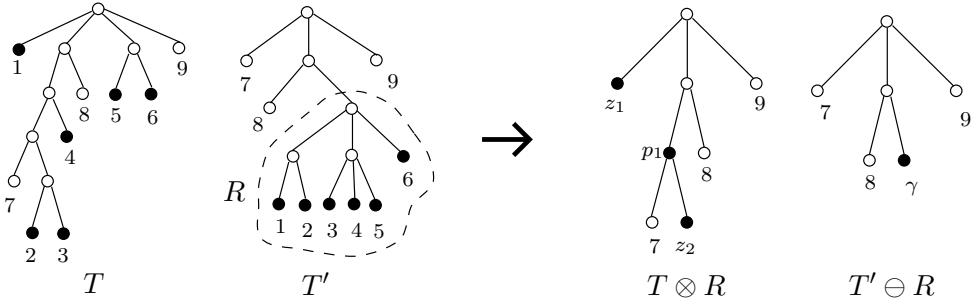}%
\end{picture}%
\setlength{\unitlength}{0.00083300in}%
\begingroup\makeatletter\ifx\SetFigFont\undefined
\def\x#1#2#3#4#5#6#7\relax{\def\x{#1#2#3#4#5#6}}%
\expandafter\x\fmtname xxxxxx\relax \def\y{splain}%
\ifx\x\y   
\gdef\SetFigFont#1#2#3{%
  \ifnum #1<17\tiny\else \ifnum #1<20\small\else
  \ifnum #1<24\normalsize\else \ifnum #1<29\large\else
  \ifnum #1<34\Large\else \ifnum #1<41\LARGE\else
     \huge\fi\fi\fi\fi\fi\fi
  \csname #3\endcsname}%
\else
\gdef\SetFigFont#1#2#3{\begingroup
  \count@#1\relax \ifnum 25<\count@\count@25\fi
  \def\x{\endgroup\@setsize\SetFigFont{#2pt}}%
  \expandafter\x
    \csname \romannumeral\the\count@ pt\expandafter\endcsname
    \csname @\romannumeral\the\count@ pt\endcsname
  \csname #3\endcsname}%
\fi
\fi\endgroup
\begin{picture}(6019,1857)(316,-1318)
\put(781,-210){\makebox(0,0)[lb]{\smash{\SetFigFont{8}{9.6}{rm}8}}}
\put(978,-212){\makebox(0,0)[lb]{\smash{\SetFigFont{8}{9.6}{rm}5}}}
\put(1207,-223){\makebox(0,0)[lb]{\smash{\SetFigFont{8}{9.6}{rm}6}}}
\put(691,-478){\makebox(0,0)[lb]{\smash{\SetFigFont{8}{9.6}{rm}4}}}
\put(316,-775){\makebox(0,0)[lb]{\smash{\SetFigFont{8}{9.6}{rm}7}}}
\put(446,-1085){\makebox(0,0)[lb]{\smash{\SetFigFont{8}{9.6}{rm}2}}}
\put(648,-1089){\makebox(0,0)[lb]{\smash{\SetFigFont{8}{9.6}{rm}3}}}
\put(1393, 80){\makebox(0,0)[lb]{\smash{\SetFigFont{8}{9.6}{rm}9}}}
\put(363, 86){\makebox(0,0)[lb]{\smash{\SetFigFont{8}{9.6}{rm}1}}}
\put(875,-1282){\makebox(0,0)[b]{\smash{\SetFigFont{11}{13.2}{rm}$T$}}}
\put(1786,-17){\makebox(0,0)[lb]{\smash{\SetFigFont{8}{9.6}{rm}7}}}
\put(1865,-271){\makebox(0,0)[lb]{\smash{\SetFigFont{8}{9.6}{rm}8}}}
\put(2707,  0){\makebox(0,0)[lb]{\smash{\SetFigFont{8}{9.6}{rm}9}}}
\put(2005,-908){\makebox(0,0)[lb]{\smash{\SetFigFont{8}{9.6}{rm}1}}}
\put(2170,-930){\makebox(0,0)[lb]{\smash{\SetFigFont{8}{9.6}{rm}2}}}
\put(2328,-930){\makebox(0,0)[lb]{\smash{\SetFigFont{8}{9.6}{rm}3}}}
\put(2478,-930){\makebox(0,0)[lb]{\smash{\SetFigFont{8}{9.6}{rm}4}}}
\put(2620,-937){\makebox(0,0)[lb]{\smash{\SetFigFont{8}{9.6}{rm}5}}}
\put(2800,-615){\makebox(0,0)[lb]{\smash{\SetFigFont{8}{9.6}{rm}6}}}
\put(1750,-706){\makebox(0,0)[b]{\smash{\SetFigFont{11}{13.2}{rm}$R$}}}
\put(2260,-1291){\makebox(0,0)[b]{\smash{\SetFigFont{11}{13.2}{rm}$T'$}}}
\put(4995,-141){\makebox(0,0)[lb]{\smash{\SetFigFont{8}{9.6}{rm}9}}}
\put(4664,-584){\makebox(0,0)[lb]{\smash{\SetFigFont{8}{9.6}{rm}8}}}
\put(4337,-1028){\makebox(0,0)[lb]{\smash{\SetFigFont{8}{9.6}{rm}7}}}
\put(4126,-130){\makebox(0,0)[b]{\smash{\SetFigFont{8}{9.6}{rm}$z_1$}}}
\put(4566,-1020){\makebox(0,0)[b]{\smash{\SetFigFont{8}{9.6}{rm}$z_2$}}}
\put(4361,-422){\makebox(0,0)[b]{\smash{\SetFigFont{8}{9.6}{rm}$p_1$}}}
\put(4571,-1291){\makebox(0,0)[b]{\smash{\SetFigFont{11}{13.2}{rm}$T \otimes R$}}}
\put(5403,-193){\makebox(0,0)[lb]{\smash{\SetFigFont{8}{9.6}{rm}7}}}
\put(5685,-633){\makebox(0,0)[lb]{\smash{\SetFigFont{8}{9.6}{rm}8}}}
\put(6279,-235){\makebox(0,0)[lb]{\smash{\SetFigFont{8}{9.6}{rm}9}}}
\put(5977,-642){\makebox(0,0)[b]{\smash{\SetFigFont{8}{9.6}{rm}$\gamma$}}}
\put(5851,-1291){\makebox(0,0)[b]{\smash{\SetFigFont{11}{13.2}{rm}$T' \ominus R$}}}
\end{picture}
\end{center}
\caption{An example of label compression.}
\label{topology}
\end{figure}

We are now ready to define the concept of label compression.
Let $T$ and $R$ be rooted evolutionary trees labeled over $L$ and $K$,
respectively.  The {\it compression} of $T$ with respect to $R$, denoted by
$\lc{T}{R}$, is a tree constructed by affixing extra nodes to $T | (L-K)$
with the following steps;
see Figure~\ref{topology} for an example.
Consider each node $y$ in $T |(L-K)$,
let $x$ be its parent in $T |(L-K)$.
\begin{itemize}
\item Let $\AAA(T,K,y)$ denote the set of subtrees of $T$ that are attached to
   $y$ and whose leaves are all labeled over $K$.
   If $\AAA(T,K,y)$ is non-empty, compress all the trees in
   $\AAA(T,K,y)$ into a single node $z_1$ and attach it to $y$.

\item Let $\HH(T,K,y)$ denote the set of subtrees of $T$ that hang between $x$
   and $y$  (by definition of $T | (L-K)$, 
   these subtrees are all labeled over $K$).
   If $\HH(T,K,y)$ is non-empty, compress the parents $p_1, \ldots,
   p_m$ of the roots of the trees in $\HH(T,K,y)$ into a single node $p_1$,
   and insert it between $x$ and $y$; 
   also compress all the trees in $\HH(T,K,y)$
   into a single node $z_2$ and attach it to $p_1$.
\end{itemize}
The nodes $z_1$, $z_2$ and $p_1$ are
called {\em compressed\/} nodes, and
the leaves in $\lc{T}{R}$ that are not
compressed are {\em atomic} leaves.

We further store in $\lc{T}{R}$ some auxiliary information about the
relationship between $T$ and $R$.  For an internal node $v$ in $\lc{T}{R}$,
let $\ralpha{}{v}$ $=$ $\MAST(T^{v},R)$.  For a compressed leaf $v$ 
in $\lc{T}{R}$,
if it is compressed from a set of subtrees $T^{v_1}, \ldots,
T^{v_s}$, let $\ralpha{}{v} = \max \{ \MAST(T^{v_1},R)$,
$\ldots, \MAST(T^{v_s},R)\}$.

Let $T_1$ and $T_2$ be two rooted evolutionary trees.
Assume $T_2$ contains a rooted subtree $R$.
Given $\lc{T_1}{R}$, we can compute $\mast(T_1,T_2)$
without examining $R$.  We first construct $\mc{T_1}{R}$ by
replacing $R$ of $T_2$ with a shrunk leaf and then
compute $\mast(T_1,T_2)$ from $\lc{T_1}{R}$ and $\mc{T_2}{R}$.  
To further our discussion, we next generalize the
definition of maximum agreement subtree for 
a pair of trees that contain compressed leaves and 
a shrunk leaf, respectively.

Let $W_1= \lc{T_1}{R}$ and $W_2 = \mc{T_2}{R}$.
Let $\cleaf$ be the shrunk leaf in $W_2$.
We define an agreement subtree of $W_1$ and $W_2$ similar to 
that of ordinary evolutionary trees.  An atomic leaf must
still be mapped to an atomic leaf with the same label.  However,
the shrunk leaf $\cleaf$ of $W_2$ can be mapped to any internal node 
or compressed leaf $v$ of $W_1$ as long as $\ralpha{}{v} > 0$.
The size of an agreement subtree is the number of its atomic leaves,
plus $\ralpha{}{v}$ if $\cleaf$ is mapped to a node $v \in W_1$.
A {\em maximum} agreement subtree of $W_1$ and $W_2$ is one with
the largest size. Let $\mast(W_1,W_2)$ denote the size of
such a subtree.  The following lemma is the cornerstone of label compression.

\begin{lemma} \label{equalmast}
$\MAST(T_1,T_2) = \mast(W_1,W_2)$.
\end{lemma}
\begin{proof}
It follows directly from the definition.
\end{proof}

We can compute $\mast(W_1,W_2)$ as if $W_1$ and $W_2$
were ordinary rooted evolutionary trees \cite{fpt95,ftfocs,Kao.Evol.Siam}
with a special procedure on handling the shrunk leaf.
The time complexity is stated in the following lemma.
Let $n = \max\{|W_1|,|W_2|\}$ and ${N} = \max\{|T_1|,|T_2|\}$.  
\begin{lemma} \label{ComputeMast}
  Suppose that all the auxiliary information of $W_1$ has
  been given.  Then
  $\mast(W_1, W_2)$ can be computed in
  $O(n^{1.5} \log {N})$
  time and afterwards we can
  retrieve $\mast(W_1^v, W_2)$ for any node $v \in W_1$ in $O(1)$ time.
\end{lemma}
\begin{proof}
We adapt Farach and Thorup's 
rooted subtree algorithm \cite{ftfocs} to compute $\mast(W_1,W_2)$.
Details are given in \S\ref{sec_app}.
\end{proof}

We demonstrate a scenario where label compression speeds up the
computation of $\mast(U_1^x,U_2)$ for Lemma~\ref{mainlemma}.  Suppose that
we can identify a rooted subtree $R$ of $U_2$ such that $x$ is mapped to a
node outside $R$, i.e., we can reduce Equation~(\ref{eqn_mast}) to
\begin{eqnarray}\label{fallincomp}
        \mast(U_1^x, U_2) & = &
        \max \{ \mast(U_1^x, U_2^y) \mid y
        \mbox{ is an internal node not in } R \}.
\end{eqnarray}
Note that every $U_2^y$ contains $R$ as a common subtree.  To avoid overlapping
computation on $R$, we construct $W = \lc{U_1^x}{R}$ and $X = \mc{U_2}{R}$.
Then $X^y = \mc{U_2^y}{R}$ and from Lemma~\ref{equalmast}, $\mast(U_1^x,
U_2^y) = $ $\mast(W, X^y)$.  We rewrite Equation~(\ref{fallincomp}) as
\begin{eqnarray} \label{newfallincomp}
  \mast(U_1^x, U_2) & = & \max\{ \mast(W, X^y) \mid y \mbox{ is an
   internal node of } X \}.
\end{eqnarray}
If $R$ is large, then $W$ and $X$ are much smaller than $U_1^x$ and $U_2$.
Consequently, it is beneficial to compress $U_1^x$ and compute
$\mast(U_1^x, U_2)$ according to Equation~(\ref{newfallincomp}).

\subsection{Label compression with respect to two rooted subtrees}
\label{section2} \label{subsec_compress_two}
Let $T$, $R_1$, $R_2$ be rooted evolutionary trees labeled over $L$, $K_1$,
$K_2$, respectively, where $K_1 \cap K_2 = \phi$.  Let $K = K_1 \cup K_2$.
The {\em compression} of $T$ with respect to $R_1$ and $R_2$, denoted by
$\lcc{T}{(R_1, R_2)}$, is a tree constructed from $T | (L - K)$ by the
following two steps.  For each node $y$ and its parent $x$ in $T | (L-K)$,
\begin{enumerate}
\item
        if $\AAA(T,K,y)$ is non-empty,
        compress all the trees in $\AAA(T,K,y)$ into a single
        leaf $z$ and attach it to $y$; create and attach an auxiliary
        node $\bar{z}$ to $y$;

\item
        if $\HH(T,K,y)$ is non-empty, compress the parents
        $p_1$, $\ldots$, $p_m$
        of the roots of the subtrees in $\HH(T,K,y)$ into a single
        node $p_1$ and insert it between $x$ and $y$;
        compress the subtrees in $\HH(T,K,y)$ into a
        single node $z$ and attach it to $p_1$;
        create and insert an auxiliary node $\bar{p}_1$
        between $p_1$ and $y$;
        create auxiliary nodes $\bar{z}$
        and $\bar{\bar{z}}$ and attach them to $p_1$ and
        $\bar{p}_1$, respectively.
\end{enumerate}
The nodes $p_1$ and $z$ are {\em compressed} nodes of $\lc{T}{(R_1,R_2)}$.
The nodes $\bar{p}_1, \bar{z}$, and $\bar{\bar{z}}$ are {\em auxiliary}
nodes. These nodes are
added to capture the topology of $T$ that is isomorphic with
the subtrees $R_1$ and $R_2$ of $T'$.

We also store auxiliary information in $\lc{T}{(R_1,R_2)}$.  Let $R^+$ be
the tree obtained by connecting $R_1$ and $R_2$ together with a node, which
becomes the root of $R^+$.

Consider the internal nodes of $\lc{T}{(R_1,R_2)}$.
If $v$ is an internal node inherited from $T | (L-K)$, then let 
$\ralpha{1}{v} = \MAST(T^v, R_1)$ and $\ralpha{2}{v} = \MAST(T^v, R_2)$.
If $p_1$ and $\bar{p}_1$ are internal nodes 
  compressed from some path $p_1, \ldots, p_m$ of $T$, 
  then only $p_1$ stores the values
  $\ralpha{1}{p_1} = \MAST(T^{p_1}, R_1)$, $\ralpha{2}{p_1} =
  \MAST(T^{p_1}, R_2)$, and $\ralpha{+}{p_1} = \MAST(T^{p_1}, R^+)$.

We do not store any auxiliary
information at the atomic leaves in
$\lc{T}{(R_1,R_2)}$.
Consider the other leaves in
$\lc{T}{(R_1,R_2)}$ based on how they are created.

{\it Case 1}:
        Nodes $z, \bar{z}$ are leaves
        created with respect to
        $\AAA(T, K, y)$ for some node $y$ in $T | (L-K)$.
        Let $\AAA(T, K, y)
        = \{T^{v_1}, \ldots, T^{v_k}\}$.
        We store the following values at $z$.

        \begin{itemize}
        \item \label{reuse}
                $\ralpha{1}{z}
                = \max \{  \MAST(T^{v_i},R_1) \mid i \in [1,k] \}$,
                $\ralpha{2}{z}
                = \max \{ \MAST(T^{v_i},R_2) \mid i \in [1,k] \}$,
                $\ralpha{+}{z}
                = \max \{ \MAST(T^{v_i},R^+) \mid i \in [1,k] \}$;
        \item
                $\rbeta{z} =
                \max \{
                \MAST(T^{v_{i}},R_1)
                + \MAST(T^{v_{i'}},R_2) \mid$
                 $T^{v_i}$ and $T^{v_{i'}}$ are
                distinct subtrees in $\AAA(T,K,y) \}$.
        \end{itemize}

{\it Case 2}:
        Nodes $z, \bar{z}$, and $\bar{\bar{z}}$ are leaves
        created with respect to the subtrees in $\HH(T,K,y)
        = \{T^{v_1}, \ldots, T^{v_k}\}$
        for some node $y$ in $T | (L-K)$.
        We store the following values at $z$:
\begin{itemize}
\item
         $\ralpha{1}{z}$,
        $\ralpha{2}{z}$,
        and $\ralpha{+}{z}$
        as in Case 1;
\item
        $\rbeta{z} = \max\{
        \MAST(T^{v_{i}},R_1) +
        \MAST(T^{v_{j}},R_2)  \mid $
        $T^{v_i}$ and $T^{v_{j}}$ are
        distinct subtrees in  $\HH(T,K,y)$
        that are attached to the same node in $T \}$;
\item
        $\rb{z} = \max\{ \MAST(T^{v_{j}},R_1)+
                \MAST(T^{v_{j'}},R_2)\mid (j,j') \in Z\}$ and

        $\rbb{z} =
         \max \{ \MAST(T^{v_{j}},R_2)+ \MAST(T^{v_{j'}},R_1)\mid
                        (j,j') \in Z\}$,

        where
        $Z$ $= \{ (j,j') \mid$
        $T^{v_j}$, $T^{v_{j'}} \in \HH(T,K,y)$
        and
        the parent of $v_j$ in $T$ is a proper ancestor of
        the parent of $v_{j'}$ \}.
\end{itemize}

Let $T_1$ and $T_2$ be rooted evolutionary trees.  Let $R_1$ and $R_2$ be
label-disjoint rooted subtrees of $T_2$.  
Let $W_1 = \lc{T}{(R_1,R_2)}$ and $W_2 = \mc{T'}{(R_1,R_2)}$.  
Below, we give the definition of a maximum agreement subtree 
of $W_1$ and $W_2$.

Let $\cleaf_1$
and $\cleaf_2$ be the two shrunk leaves in $W_2$ representing $R_1$
and $R_2$, respectively.  
Let $y_c$ be the least common ancestor of  $\cleaf_1$
and $\cleaf_2$ in $W_2$.
Intuitively, in a pair of agreement subtrees $(W_1',W_2')$ of $W_1$ and $W_2$,
atomic leaves are mapped to atomic leaves, and 
shrunk leaves are mapped to internal nodes or leaves.
Moreover, we allow $W_2'$ to contain $y_c$ as a leaf, which can be
mapped to an internal node or leaf of $W_1'$. More formally,
we require that there is an
isomorphism between $W_1'$ and $W_2'$ satisfying the following conditions:
\begin{enumerate} 
\item Every atomic leaf is mapped to an atomic leaf with the same label.
\item If $W_2'$
contains $y_c$ as a leaf and thus neither $\cleaf_1$ nor $\cleaf_2$ is
found in $W_2'$, then $y_c$ is mapped to a node $v$ with
$\ralpha{+}{v} > 0$.  
\item If only one of $\cleaf_1$ and $\cleaf_2$
exists in $W_2'$, say $\cleaf_1$, then it is mapped to a node
$v$ with $\ralpha{1}{v} > 0$.  
\item If both $\cleaf_1$ and $\cleaf_2$
exist in $W_2'$, then any of the following cases is permitted:
\begin{itemize} 
\item $\cleaf_1$ and $\cleaf_2$ are respectively
mapped to a compressed leaf $z$ and its sibling $\bar{z}$ in $W_1'$
with $\rbeta{z} > 0$. 
\item $\cleaf_1$ and $\cleaf_2$ are respectively
mapped to a compressed leaf $z$ and the accompanying auxiliary leaf
$\bar{\bar{z}}$ in $W_1'$ with $\rb{z} > 0$, or the leaves
$\bar{\bar{z}}$ and $z$ in $W_1'$ with $\rbb{z} > 0$. 
\item $\cleaf_1$
and $\cleaf_2$ are respectively mapped to two leaves or internal nodes
$v$ and $w$ with $\ralpha{1}{v}$, $\ralpha{2}{w} > 0$. 
\end{itemize}
\end{enumerate} 
The way we measure the size of $W_1'$ and $W_2'$ depends on
their isomorphism.
For example, if $y_c$ is mapped to some node $v$ in $W_1'$,
then the size is the total
number of atomic leaves in $W_1'$ plus $\ralpha{+}{v}$.
More precisely, the size of $W_1'$ and $W_2'$ is defined
to be the total number of atomic leaves in $W'_1$ plus
the corresponding $\alpha$ or $\beta$ values depending
on the isomerphism between $W'_1$ and $W'_2$.
A {\it maximum} agreement subtree of $W_1$ and $W_2$
is one with the largest possible size.
Let $\mast(W_1,W_2)$ denote the size of such a subtree.
The following lemma, like Lemma~\ref{equalmast}, is also
the cornerstone of label compression.

\begin{lemma}\label{lemma_ww_two}
  $\MAST(T_1, T_2)=\mast(W_1, W_2)$.
\end{lemma}
\begin{proof}
It follows directly from the definition of $\mast(W_1,W_2)$.
\end{proof}

Again, $\mast(W_1,W_2)$ can be computed by adapting Farach and Thorup's
rooted subtree algorithm \cite{ftfocs}.
The time complexity is stated in the following lemma.
Let $n$ $=$ $\max\{|W_1|$,$|W_2|\}$ and ${N} =
\max\{|T_1|,|T_2|\}$.

\begin{lemma} \label{lcclemma}
  Suppose that all the auxiliary information of $W_1$ has
  been given.  Then  we can compute $\mast(W_1,W_2)$ in        
  $O(n^{1.5} \log {N})$ time.  Afterwards we
  can retrieve $\mast(W_1^v, W_2)$ for any $v \in W$ in $O(1)$ time.
\end{lemma}
\begin{proof}
  See \S\ref{sec_app}.
\end{proof}

\section{Computing \mathversion{bold}\kmast$(U_1^x, U_2)$ --- Proof
of Lemma~\protect{\ref{mainlemma}}}
\label{sec_mast_u1x_u2}
At a high level, we first apply label compression to the input instance
$(U_1^x, U_2)$.  We then reduce the problem to a number of smaller
subproblems $(W,X)$, each of which is similar to $(U_1^x, U_2)$ and is
solved recursively.  For each $(W,X)$ generated, $X$ is a subtree of $U_2$
with at most two shrunk leaves, and $W$ is a label compression of $U_1^x$
with respect to some rooted subtrees of $U_2$ that are represented by the
shrunk leaves of $X$.  Also, $W$ and $X$ contain the same number of
atomic leaves.

\subsection{Recursive computation of \mathversion{bold}\kmast$(W,X)$}
\label{subsec_recur_mast_w_x}
Our subtree algorithm
initially sets
$W= U_1^x$ and $X= U_2$. In general,
$W = \lc{U_1^x}{R}$ and $X = \mc{U_2}{R}$,
or $W = \lc{U_1^x}{(R,R')}$ and $X = \mc{U_2}{(R,R')}$ for
some rooted subtrees $R$ and $R'$ of $U_2$.
If $W$ or $X$ has at most two nodes,
then $\mast(W,X)$ can easily be computed in linear time.  Otherwise, both
$W$ and $X$ each have at least three nodes.
Let $N = \max \{ |U_1|, |U_2| \}$ and
$n = \max \{|W|, |X|\}$.
Our algorithm first finds a
separator $y$ of $X$ and computes $\mast(W,X)$ for the following two cases.
The output is the larger of the two cases.  Figure~\ref{algo2} outlines our
algorithm.
\begin{figure}[t]
\begin{center}
\fbox{%
\small
\begin{minipage}{\mini}
\begin{list}{}{\setlength{\leftmargin}{0.5em}}
\item
/* $W$ is a rooted tree with compressed leaves.  $X$ is unrooted
with shrunk leaves. */
\item {$\mast(W, X)$}
 \begin{list}{}{\setlength{\leftmargin}{1em}}
  \item let $y$ be a separator of $X$;
  \item $\kval = \mast(W, X^y)$;
  \item {if} ($X$ has at most one shrunk leaf) {or}
   ($y$ lies between the two shrunk leaves)
  \item[] {then}
  \begin{list}{}{\setlength{\leftmargin}{1em}}
   \item {new\_subproblem}$(W, X, y)$;
   \item {for each} $(W_i, X_i)$, $\kval_i = \mast(W_i, X_i)$;
  \end{list}
  \item[] {else}
  \begin{list}{}{\setlength{\leftmargin}{1em}}
  \item let $y'$ be the node on the path between the
      two shrunk leaves that is the closest to $y$;
  \item $\kval = \mast(W, X^{y'})$;
  \item {new\_subproblem}$(W, X, y')$;
  \item {for each} $(W_i, X_i)$, set $\kval_i = \mast(W_i, X_i)$;
  \end{list}
  \item return $\max \{ \kval, \max_{i=1}^b \kval_i \}$;
 \end{list}
\end{list}

\

\begin{list}{}{\setlength{\leftmargin}{0.5em}}
\item
/* Generate new subproblems $\{ (W_1, X_1), \ldots, (W_b, X_b) \}$. */
\item {new\_subproblem}$(W, X, y)$
\begin{list}{}{\setlength{\leftmargin}{1em}}
\item let $v_1, \ldots, v_b$ be the neighbors of $y$ in $X$;
\item {for all} $i \in [1,b]$
\begin{list}{}{\setlength{\leftmargin}{1em}}
\item let $X_i$ be the unrooted tree formed by shrinking
      the subtree $X^{v_iy}$ into a shrunk leaf;
\item let $W_i$ be the rooted tree formed by compressing
      $W$ with respect to $X^{v_iy}$;
\end{list}
\item compute and store the auxiliary information in $W_i$
   for all $i \in [1,b]$;
\end{list}
\end{list}
\end{minipage}
}
\end{center}
\caption{Algorithm for computing $\mast(W, X)$.\label{algo2}}
\end{figure}

{\it Case 1}: $\mast(W,X) = \mast(W, X^y)$.  We root $X$ at
$y$ and evaluate $\mast(W,X^y)$.  By Lemma~\ref{lcclemma}, this takes
$O(n^{1.5} \log N)$
time.

{\it Case 2}: $\mast(W,X) = \mast(W,X^z)$ for some internal
node $z \neq y$.  We compute
$\max \{ \mast(W,X^z) \mid z \mbox{ is an internal
node and } z \neq y \}$ by solving a set of subproblems $\{
\mast(W_1,X_1)$, $\ldots$, $\mast(W_b,X_b)\}$ where
their total size is $n$ and $\max \{ \mast(W,X^z) \mid z
\mbox{ is an internal node and } z \neq y \} =$ $\max
\{ \mast(W_i,X_i) \mid i \in [1,b] \}$.  Moreover, our algorithm enforces
the following properties.
\begin{itemize}
\item If $X$ contains at most one shrunk leaf, every subproblem generated
  has
size at most half that of $X$.
\item If $X$ has two shrunk leaves,
at most one subproblem
$(W_{i_o}, X_{i_o})$ has size
greater than half that of $X$,
but $X_{i_o}$  contains only one shrunk leaf.
Thus, in the next recursion level,
 every subproblem spawned by $(W_{i_o}, X_{i_o})$
has size at most half that of $X$.
\end{itemize}
To summarize, whenever the recursion gets down by two levels,
the size of a subproblem reduces by half.

The subproblems
$\mast(W_1,X_1), \ldots$,
$\mast(W_b,X_b)$
are formally defined as follows.
Assume that the separator $y$  has $b$ neighbors in $X$, namely,
$v_1, \ldots, v_b$.
For each $i \in [1,b]$,
let $C_i$ be the connected component in $X - \{y\}$ that
contains $v_i$.
        The size of $C_i$ is at most half that of
        $X$.
Intuitively, we would like to shrink the subtree
        $X^{v_iy}$ into a leaf, producing a smaller unrooted tree $X_i$.
We
first consider the simple case where $X$ has at most one shrunk leaf.
Then no
$C_i$ contains more than one shrunk leaf.

If $C_i$ contains no shrunk leaf, then
        $X_i$ contains only one shrunk leaf
        representing the subtree $X^{v_iy}$.
        Note that $X^{v_iy}$ corresponds to the subtree
        $U_2^{v_iy}$ in $U_2$ and $X_i
          = \mc{U_2}{U_2^{v_iy}}$.
Let $W_i=\lc{U_1^x}{U_2^{v_iy}}$.

If $C_i$ contains one shrunk leaf
 $\cleaf_1$ 
 then $X_i$ contains $\cleaf_1$ as well
        as a new shrunk leaf
        representing the subtrees $X^{v_iy}$.
        The two subtrees are label-disjoint.
        Again, $X^{v_iy}$ corresponds to the subtree $U_2^{v_iy}$
        in $U_2$.  Assume that $\cleaf_1$
        corresponds to a subtree $U_2^{v'y'}$ in $U_2$. Then
                $X_i= \mc{U_2}{(U_2^{v'y'}, U_2^{v_i y})}$.
        Let $W_i = \lcc{U_1^x}{(U_2^{v'y'}, U_2^{v_i y})}$.

We now consider the case
where $X$ itself already has two shrunk leaves $\cleaf_1$ and $\cleaf_2$.
If $y$ lies on the path between $\cleaf_1$ and $\cleaf_2$,
        then no $C_i$ contains more than one shrunk leaf
        and we define the smaller problem instances $(W_i,X_i)$ as above.
Otherwise, there is a $C_i$
        containing both $\cleaf_1$ and $\cleaf_2$.
$X_i$ as defined contains three compressed leaves,
violating our requirement.
In this case, we replace $y$
with the node $y'$ on the path between
$\cleaf_1$ and $\cleaf_2$, which is the closest
to $y$.
Now, to compute $\mast(W, X)$, we consider the two cases
depending on whether the root of $W$
is mapped to $y'$ or not.  Again,
we first compute $\mast(W,X^{y'})$. Then,
we define the connected components
$C_i$ and the smaller problem instances $(W_i,X_i)$
with respect to $y'$.
Every $X_i$  has at most two compressed leaves,
but $y'$ may not be a separator and we cannot guarantee that
        the size of every subproblem is reduced by half.
However,
there can exist  only one connected component $C_{i_o}$
         with size larger than half that of $X$.
Indeed, $C_{i_o}$ is the component containing $y$.
In this case,
        both $\cleaf_1$ and $\cleaf_2$ are not inside $C_{i_o}$,
        and $X_{i_o}$ as defined
        contains only one compressed leaf.
Thus, the subproblems that $\mast(W_{i_0},X_{i_0})$ spawns
in the next recursion level
each have size of at most half that of $(W,X)$.

With respect to $y$ or $y'$, computing the topology of all $X_i$ and
$W_i$ from $X$ and $W$ is straightforward; see
\S\ref{topologySection}.  Computing the auxiliary information in all
$W_i$ efficiently requires some intricate techniques, which are
detailed in \S\ref{sec_aux_0} and \S\ref{sec_aux_1_2}.

\subsection{Computing the topology of compressed trees}
\label{topologySection}
The topology of $X_i$ can be constructed from $X$ by
replacing
the subtree $X^{v_iy}$ of $X$ with a shrunk leaf.
Let $J$ and $J_i$ be the sets of labels in
$X$ and $X_i$, respectively.
For the trees $W_i$, recall that
the definitions of $W$ and the trees $W_i$ are based
on affixing some nodes to
the trees $U_1^x | J$ and $U_1^x| J_i$, respectively.
Observe that
$W | J$ and $ U_1^x | J$
have the same topology.
Moreover,
$W | J_i  = (W|J)|J_i$  and
$U_1^x | J_i = (U_1^x | J) | J_i$.
Thus,
$W | J_i$  and
$U_1^x | J_i$
have the same topology.
We can obtain  $U_1^x|J_i$ by constructing $W|J_i$.
Note that
$J = \bigcup_{1 \leq i \leq b} J_i$ and
all the label sets $J_i$ are disjoint.
We can construct all the trees $W | J_i$ from $W$
in $O(n)$ time \cite{ch96,ft95.infcomp}.
Next, we show how to
construct $W_i$ from $W | J_i$
in time linear in the size of $W|J_i$.
We only detail the case where $X_i$ consists of two
shrunk leaves.
The case for
one shrunk leaf is similar.
The following procedure is  derived directly from
the definition of the compression of $U_1^x$ with respect
to two subtrees.

Let $v$ be any node of $W | J_i$. If $v$ is not the root,
let $u$ be the parent of $v$ in $W | J_i$.
\begin{itemize}
\item
If $\AAA(U_1^x, L-J_i, v)$ is non-empty or equivalently
the degree of $v$ in $U_1^x$ is different from its degree
in $W | J_i$,  then
attach auxiliary leaves $z$ and $\bar{z}$ to $v$.
\item
If $\HH(U_1^x, L-J_i, v)$ is non-empty or equivalently
$u$ is not the parent of $v$ in $U_1^x$, then
create a path between $u$ and $v$
consisting of two nodes
$p$ and $\bar{p}$, attach auxiliary leaves $z$ and $\bar{z}$
to $p$, and attach $\bar{\bar{z}}$ to $\bar{p}$.
\end{itemize}

\subsection{Time complexity of computing \mathversion{bold}\kmast($W, X$)}\

\begin{lemma}
  We can compute $\mast(W, X)$ in
$O(n^{1.5} \log N)$
time.
\end{lemma}
\begin{proof}
  Let $T(n)$ be the computation time of $\mast(W, X)$.  The computation is
  divided into two cases.  Case 1 of \S\ref{subsec_recur_mast_w_x} takes
$O(n^{1.5} \log N)$
time.
  For Case 2, a set of subproblems $\{ \mast(W_i,
  X_i) \mid i \in [1, b] \}$ are generated.  As to be shown in
  $\S$\ref{sec_aux_0} and $\S$\ref{sec_aux_1_2},
  the time to prepare all these
  subproblems is also
$O(n^{1.5} \log N)$.
  These subproblems, except
  possibly one, are each of size less than $n/2$.  For the exceptional
  subproblem, say, $\mast(W_l, X_l)$, its computation is again divided into
  two cases.  One case takes
$O(n^{1.5} \log N)$
time.  For the other case,
  another set of subproblems are generated in
$O(n^{1.5} \log N)$
time.  This
  time every such subproblem has size less than $n/2$.  Let
  $\Sigma$ be
  the set of all the subproblems generated in both steps.  The total size
  of the subproblems in $\Sigma$ is at most $n$, and
\[
T(n) = O(n^{1.5} \log N)
+ \sum_{\mast(W', X') \in \Sigma} T(|X'|).
\]
It follows that $T(n) =$
$O(n^{1.5} \log N)$.
\end{proof}

By letting $W = U_1^x$ and $X = U_2$, we have proved Lemma~\ref{mainlemma}.
What remains is to show how to compute the auxiliary information stored in
all $W_i$ from $(W,X)$ in
$O(n^{1.5} \log N)$
time.  Note that $X$
contains at most two shrunk leaves.  Depending on the number of shrunk
leaves in $X$, we divide our discussion into \S\ref{sec_aux_0} and
\S\ref{sec_aux_1_2}.

\section{Auxiliary information for \mathversion{bold}$X$ with no shrunk leaf}
\label{sec_aux_0}
The case of $X$ containing no shrunk leaf occurs only when the
algorithm starts, i.e., $W= U_1^x$, $X=U_2$. and $N=n$.
The subproblems
$\mast(W_1,X_1),$
$\ldots$, $\mast(W_b,X_b)$ spawned from $(W,X)$
are defined by an internal node $y$ in $X$,
which is adjacent to the nodes $v_1, \ldots, v_b$.
Let $R_i$ and $\complement{R}_i$ denote the rooted subtrees $X^{v_iy}$
and $X^{yv_i}$, respectively.
Note that
  the rooted tree $X^y$
  is composed of the subtrees $\complement{R}_1,
  \ldots, \complement{R}_b$. Also, $W_i = \lc{W}{R_i}$
  and $X_i = \mc{W}{R_i}$.
\begin{figure}
\label{struct}
\begin{center}
\begin{picture}(0,0)%
\epsfig{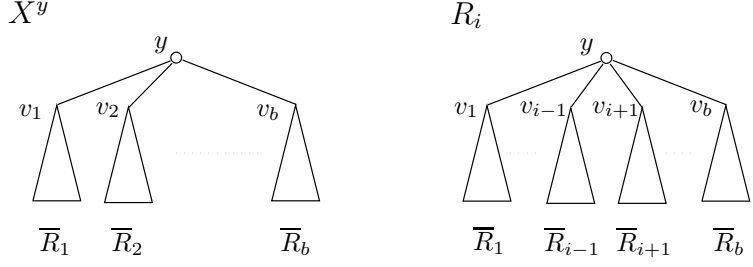}%
\end{picture}%
\setlength{\unitlength}{0.00083300in}%
\begingroup\makeatletter\ifx\SetFigFont\undefined
\def\x#1#2#3#4#5#6#7\relax{\def\x{#1#2#3#4#5#6}}%
\expandafter\x\fmtname xxxxxx\relax \def\y{splain}%
\ifx\x\y   
\gdef\SetFigFont#1#2#3{%
  \ifnum #1<17\tiny\else \ifnum #1<20\small\else
  \ifnum #1<24\normalsize\else \ifnum #1<29\large\else
  \ifnum #1<34\Large\else \ifnum #1<41\LARGE\else
     \huge\fi\fi\fi\fi\fi\fi
  \csname #3\endcsname}%
\else
\gdef\SetFigFont#1#2#3{\begingroup
  \count@#1\relax \ifnum 25<\count@\count@25\fi
  \def\x{\endgroup\@setsize\SetFigFont{#2pt}}%
  \expandafter\x
    \csname \romannumeral\the\count@ pt\expandafter\endcsname
    \csname @\romannumeral\the\count@ pt\endcsname
  \csname #3\endcsname}%
\fi
\fi\endgroup
\begin{picture}(5309,1616)(197,-3096)
\put(4066,-1808){\makebox(0,0)[b]{\smash{\SetFigFont{10}{12.0}{rm}$y$}}}
\put(1410,-1793){\makebox(0,0)[b]{\smash{\SetFigFont{10}{12.0}{rm}$y$}}}
\put(2077,-2220){\makebox(0,0)[b]{\smash{\SetFigFont{10}{12.0}{rm}$v_b$}}}
\put(600,-2217){\makebox(0,0)[b]{\smash{\SetFigFont{10}{12.0}{rm}$v_1$}}}
\put(1080,-2205){\makebox(0,0)[b]{\smash{\SetFigFont{10}{12.0}{rm}$v_2$}}}
\put(4780,-2194){\makebox(0,0)[b]{\smash{\SetFigFont{10}{12.0}{rm}$v_b$}}}
\put(3314,-2204){\makebox(0,0)[b]{\smash{\SetFigFont{10}{12.0}{rm}$v_1$}}}
\put(3798,-2200){\makebox(0,0)[b]{\smash{\SetFigFont{10}{12.0}{rm}$v_{i-1}$}}}
\put(4250,-2200){\makebox(0,0)[b]{\smash{\SetFigFont{10}{12.0}{rm}$v_{i+1}$}}}
\put(751,-3061){\makebox(0,0)[b]{\smash{\SetFigFont{10}{12.0}{rm}$\overline{R}_1$}}}
\put(1201,-3061){\makebox(0,0)[b]{\smash{\SetFigFont{10}{12.0}{rm}$\overline{R}_2$}}}
\put(2251,-3061){\makebox(0,0)[b]{\smash{\SetFigFont{10}{12.0}{rm}$\overline{R}_b$}}}
\put(3976,-3061){\makebox(0,0)[b]{\smash{\SetFigFont{10}{12.0}{rm}$\overline{R}_{i-1}$}}}
\put(3458,-3058){\makebox(0,0)[b]{\smash{\SetFigFont{10}{12.0}{rm}$\overline{R}_1$}}}
\put(4426,-3061){\makebox(0,0)[b]{\smash{\SetFigFont{10}{12.0}{rm}$\overline{R}_{i+1}$}}}
\put(4953,-3063){\makebox(0,0)[b]{\smash{\SetFigFont{10}{12.0}{rm}$\overline{R}_b$}}}
\put(3316,-1643){\makebox(0,0)[b]{\smash{\SetFigFont{12}{14.4}{rm}$R_i$}}}
\put(586,-1636){\makebox(0,0)[b]{\smash{\SetFigFont{12}{14.4}{rm}$X^y$}}}
\end{picture}
\end{center}
\caption{The structures of $X^y$ and $R_i$. \label{shrunk_tree}}
\end{figure}
  The total size of all $\complement{R}_i$ is
        at most $n$.
        Furthermore, each $R_i$ is
$X^{y}$ with $\complement{R}_i$ removed; see Figure~\ref{shrunk_tree}.
This section discusses how to
compute
the auxiliary information required by each $W_i$ in
$O(n^{1.5} \log N)$
time.

\subsection{Auxiliary information in the compressed leaves of
  \mathversion{bold}$W_i$}
\label{SectionCase1-leaf}
Consider any compressed leaf $v$ in $W_i$.  Let $S_v$ denote the set of
subtrees from which $v$ is compressed.  Then, the auxiliary information to
be stored in $v$ is
\begin{equation}\label{auxinfo}
        \ralpha{}{v} = \max \{ \mast(W^z, R_i) \mid W^z \in S_v \}.
\end{equation}
Observe that for any $W^z \in S_v$,
$W^z$ contains no labels outside $R_i$.
So $\mast(W^z, R_i) =$ $\mast(W^z, X^y)$ and we can rewrite
Equation~(\ref{auxinfo}) as
\[
        \ralpha{}{v}=\max\{ \mast(W^{z},X^y) \mid W^z \in S_v \}.
\]
We use the rooted subtree algorithm of \cite{ftfocs} to compute
$\mast(W,X^y)$ in $O(n^{1.5} \log N)$ time.  Then, we can retrieve the
value of $\mast(W^z, X^y)$ for any node $z \in W$ in $O(1)$ time.  To
compute $\max \{ \mast(W^z, X^y) \mid W^z \in S_v \}$ efficiently, we
assume that for any node $u \in W$, the subtrees attached to $u$ are
numbered consecutively, starting from 1.  We consider a preprocessing
for efficient retrieval of the following types of values:
\begin{itemize}
\item for some node $u \in W$ and some interval $[a,b]$,
        $\max\{ \mast(W^z,X^y) \mid W^z$
        is a subtree attached
                to $u$ and its number
                falls in $[a,b] \}$;
\item
        for some path $P$ of $W$,
            $\max \{ \mast(W^z, X^y) \mid
         W^z$ is a subtree
        attached to some node in $ P \}$.
\end{itemize}

\begin{lemma} \label{productquery}
  Assume that we can retrieve $\mast(W^z,X^y)$ for any
  $z \in W$ in $O(1)$
  time. Then we can preprocess $W$ and $X$ and construct additional data
  structures in $O(n\log^* n)$ time so that any value of the above types
  can be retrieved in $O(1)$ time.
\end{lemma}
\begin{proof}
  We adapt preprocessing techniques for on-line product queries in
  \cite{Noba}.
\end{proof}

With the preprocessing stated in Lemma~\ref{productquery}, we
can determine
$\ralpha{}{v}$ as follows.
Note that $S_v$
is either
a subset of the
subtrees
attached to a node $u$ in $W$
or
 the set of subtrees
attached to
  nodes on a particular path in $W$.
In the former case,
        $u$ is also a parent of $v$ and
        $S_v$ is partitioned into at most $d_u +1$ intervals
        where $d_u$ is the degree of $u$ in $W_i$.
        {From} Lemma~\ref{productquery},
        $\ralpha{}{v}$ can be found
                in $O(d_u + 1)$ time.
Similarly, for the latter case,
        $\ralpha{}{v}$ can be found in $O(1)$ time.
Thus,
the compressed leaves
in $W_i$ are processed in
$O(|W_i|)$ time.
Summing over
all $W_i$, the time complexity is
$O( n )$.
Therefore, the overall computation time for
 preprocessing and
 finding auxiliary information
 in the leaves of all $W_i$ is
$O(n^{1.5} \log N)$.

\subsection{Auxiliary information in the internal
nodes of \mathversion{bold}${W_i}$} Consider any internal node $v$ in
$W_i$ with $i \in [1,b]$.  Our goal is to compute the auxiliary
information $\ralpha{}{v}=\mast(W^v,R_i)$.  Note that $R_i$ may be of
size $\Theta(n)$, and even computing one particular $\mast(W^v,R_i)$
already takes $O(n^{1.5}\log N)$ time.  Fortunately, these $R_i$ are
very similar. Each $R_i$ is $X^y$ with $\complement{R}_i$ removed.
Exploiting this similarity and using the algorithm in
\S\ref{all_cavity} for all-cavity matchings, we can perform an
$O(n^{1.5} \log N)$-time preprocessing so that we can retrieve
$\mast(W^v, R_i)$ for any internal node $v$ in $W$ and $i \in [1,b]$
in $ O(\log^2 n)$ time.  Therefore, it takes $O(|W_i| \log^2 n)$ time
to compute $\ralpha{}{v}$ for all internal nodes $v$ of one particular
$W_i$, and $O(n \log^2 n)$ time for all $W_i$.  The $O(n^{1.5} \log
N)$-time preprocessing is detailed as follows.

First, note that if we remove
 $y$ from $X^y$, the tree would decompose
into    the subtrees $\complement{R}_1,
  \ldots, \complement{R}_b$.
Thus,
  the total size of all $\complement{R}_i$ is
  at most $n$.
The next lemma suggests a way
to retrieve efficiently $\mast(W^v, \complement{R}_i)$
and $\max \{ \mast(W^v, \complement{R}_j) \mid j \in I \}$
for any $v \in W$ and and $I \subseteq [1, b]$.
\begin{lemma}
\label{EasyValues}
        We can compute
        $\mast(W, \complement{R}_i)$
        for all $i \in [1,b]$ in
$O(n^{1.5} \log N)$
        time.
Then, we can retrieve
         $\mast(W^v, \complement{R}_i)$
        for any node $v$ in $W$ and $i \in [1,b]$
        in $O(\log n)$ time.
Furthermore, we can build a data structure
to retrieve
$\max \{ \mast(W^v, \complement{R}_j) \mid j \in I \}$
for any $v \in W$ and $I \subseteq [1, b]$
in $O(\log^2 n)$ time.
\end{lemma}
\begin{proof}
This lemma follows from the rooted subtree algorithm and related data
structures in \cite{ftfocs}.
\end{proof}

Below, we give a formula
to compute $\mast(W^v, R_i)$
efficiently.
For any $z \in W$ and $i \in [1,b]$, let
$\rrmast(W^z, R_i)$ denote the maximum size among all the
agreement subtrees of $W^z$ and $R_i$
in which $z$ is
mapped to the root of $R_i$.
\begin{lemma}
\label{abcd}
\[\displaystyle
\mast(W^v, R_{{i}})
=
\max \left\{
        \begin{array}{l}
        \max \{ \mast(W^v, \complement{R}_j) \mid j \in [1,b],
               j \neq i\}; \\
        \max \{ \rrmast(W^z, R_i) \mid
                z \in W^v \}. \\
        \end{array}
\right.
\]
\end{lemma}
\begin{proof}
Observe that
$\mast(W^v,R_i) = \mast(W^z,R_i) = \rrmast(W^z,R_i)$ if
in some maximum agreement subtree of $W^v$ and $R_i$,
the root of $R_i$ is mapped to some node $z$ in $W^v$.
On the other hand,
         $\mast(W^v,R_i) = \mast(W^v,\complement{R}_j)$
         for some $j \neq i$
if
        in some maximum agreement subtree of $W^v$ and $R_i$,
         the root of $R_i$ is not mapped to any node $z$ in $W^v$.
\end{proof}

By Lemma~\ref{abcd}, we decompose the computation of $\mast(W^v,R_i)$
into two parts.  The value $\max\{ \mast(W^v, \complement{R}_j) \mid
j\in[1,b], j\neq i\}$ is determined by answering two queries $\max \{
\mast(W^v, \overline{R_j}) \mid j \in [1, i-1] \}$ and $\max \{
\mast(W^v, \overline{R_j}) \mid j \in [i+1, b] \}$ in $O(\log^2 n)$
time by Lemma~\ref{EasyValues}.  The computation of $ \max \{
\rrmast(W^z, R_i) \mid z \in W^v \}$ makes use of a maximum weight
matching of some bipartite graph as follows.

Let $\children(z)$ denote the set of children of a node $z$ in a
tree. Let $G_{z,i} \subseteq \children(z) \times \{\complement{\nR}_1,
\ldots, \complement{\nR}_{i-1}, \complement{\nR}_{i+1}$, $\ldots,
\complement{\nR}_b \}$ be a bipartite graph where $w \in \children(z)$
is connected to $\complement{\nR}_j$ if and only if $\mast(W^{w},
\complement{R}_j) > 0$.  Such an edge has weight
$\mast(W^{w},\complement{R}_j) \leq N$.

\begin{fact}[see~\cite{ftfocs}]
If the root of $R_i$ is mapped to $z$ in some maximum agreement
subtree of $W^z$ and $R_i$, then a maximum weight matching of
$G_{z,i}$ consists of at least two edges, and $\mwm(G_{z,i}) =
\rrmast(W^z,R_i)$.
\end{fact}

Note that if a maximum weight matching of $G_{z,i}$ consists of one
edge, it corresponds to an agreement subtree of $W^z$ and $R_i$ in
which the root of $R_i$ is not mapped to any node in $W^z$.  Thus, it
is possible that $\mwm(G_{z,i}) > \rrmast(W^z,R_i)$.  Nevertheless, in
this case we are no longer interested in the exact value of
$\rrmast(W^z,R_i)$ since in a maximum agreement subtree of $W^z$ and
$R_i$, the root of $R_i$ is not mapped to any node in $W^z$.  In fact,
Lemma~\ref{abcd} can be rewritten with the $\rrmast(W^z,R_i)$ replaced
by $\mwm(G_{z,i})$.  Furthermore, since $G_{z,1}$, $G_{z,2}$,
$\ldots$, $G_{z,b}$ are very similar, the weights of a maximum weight
matching cannot be all distinct.
\begin{lemma} \label{sparse}
  At least $b-d_z$ of
  $\mwm(G_{z,1})$, $\mwm(G_{z,2})$, $\ldots$, $\mwm(G_{z,b})$
  have the same value,
  where $d_z$ denotes the degree of $z$ in $W$.
\end{lemma}
\begin{proof}
Consider the bipartite graph ${K} \subseteq
\children(z) \times
\{ \complement{R}_1,
\ldots, \complement{R}_b \}$
in which a node $w \in \children(z)$ is connected to
$\complement{R}_i$ if and only
if $\mast(W^w, \complement{R}_i) > 0$. This edge
is given a weight of $\mast(W^w,\complement{R}_i)$.
Then, every $G_{z,i}$ is a subgraph of ${K}$.
Let $M$ be a maximum weight matching of ${K}$.
Observe that if an $R_i$ is not adjacent to
any edge in $M$, then $M$ is also a maximum weight
matching of $G_{z,i}$.
Since $M$ contains at most $d_z$ edges,
there are at least $b-d_z$ trees $R_i$ not
adjacent to any edge in $M$ and the corresponding
$\mwm(G_{z,i})$ have the same value.
\end{proof}

We next use $O(n^{1.5} \log N)$ time to find for all $z$ in $W$,
$\mwm(G_{z,1}),\ldots,\mwm(G_{z,b})$.  The results are to be stored in
an array $A_z$ of dimension $b$ for each node $z$, i.e.,\ $A_z[i] =
\mwm(G_{z,i})$.  Note that if we represent each $A_z$ as an ordinary
array, then filling these arrays entry by entry for all $z \in W$
would cost $\Omega(bn)$ time.  Nevertheless, by Lemma~\ref{sparse},
most of the weights $\mwm(G_{z,i})$ have the same value.  Thus, we
store these values in sparse arrays.  Like an ordinary array, any
entry in a {\em sparse array} $A$ can be read and modified in $O(1)$
time.  In addition, we require that all the entries in $A$ can be
initialized to a fixed value in $O(1)$ time and that all the distinct
values stored in $A$ can be retrieved in $O(m)$ time, where $m$
denotes the number of distinct values in $A$.  For an implementation
of sparse array, see Exercise 2.12, page 71 of \cite{AHU}.

Before showing how to build these sparse arrays, we illustrate how
they support the computation of
\begin{equation} \label{online}
        \max \{ \mwm(G_{z,i}) \mid
                \mbox{$z \in W^v$}\} =
        \max \{ A_z[i] \mid
                \mbox{$z \in W^v$}\}.
\end{equation}
An efficient data structure for answering such a query is given in
\S\ref{app_b}.
Let $m_z$ be the number of distinct values in $A_z$, and $m = \sum_{z
\in W} (m_z + 1)$.  Let $\alpha(n)$ denote the inverse Ackermann
function.  Appendix \ref{app_b} shows how to construct a data
structure on top of the sparse arrays $A_z$ in $O(m \alpha(|W|))$
time such that we can retrieve for any $v \in W$ and $i \in [1,b]$ the
value of $\max \{ A_z[i] \mid z \in W^v \}$ in $O(\log |W| )$ time.
{From} Lemma~\ref{sparse}, $m_z \leq d_z +1$ for all $z \in W$; thus,
$m = O(|W|)$.  Therefore, the data structure can be built in time
$O(m\alpha(|W|)) = O(|W|\alpha(|W|)) = O(n \log n)$ and the retrieval
time of Equation~(\ref{online}) is $O(\log |W|) = O(\log n)$.

To summarize, after
building all the necessary data structures,
we can retrieve
        $\max \{ \mast(W^v, \complement{R}_j) \mid j \in [1,b],
               j \neq i\}$
in $O(\log^2 n)$ time
and
        $\max \{ \rrmast(W^z, R_i) \mid z \in W^v \}$
in $O(\log n )$ time.
Hence,
for any $v \in W$ and
$i \in [1,b]$,
$\mast(W^v, R_i)$
can be computed in $O(\log^2 n)$ time.

To complete our discussion,
we show below how to construct a sparse array $A_z$
or equivalently compute the weights
$\{\mwm(G_{z,i}) \mid i \in [1,b]\}$
efficiently. We cannot afford to examine every $G_{z,i}$ and compute
$\mwm(G_{z,i})$ separately.
Instead we build only one
 weighted graph
${\Gz}
 \subseteq \children(z) \times
 \{\complement{\nR}_1, \ldots, \complement{\nR}_b \}$
as follows.

For a node $z$ in $W$,
        the  {\em max-child} $z'$ of $z$ is a child of $z$ such that
        the subtree rooted at $z'$
        contains the maximum number of atomic leaves
        among all the subtrees attached to $z$.
Let $\kappa(z)$ denote the total number of atomic leaves that
are in $W^z$ but not in $W^{z'}$.
The edges of $\Gz$ are specified as follows.
\begin{itemize}
\item
For any non-max-child $u$ of $z$,
$\Gz$ contains an edge between $u$ and some $\complement{\nR}_i$
if and only if $\mast(W^{u}, \complement{R}_i) > 0$.
There are at most $\kappa(z)$ such edges.

\item
Regarding the max-child $z'$ of  $z$,
we only put into $\Gz$ a limited number of edges between
        $z'$ and $\{\complement{\nR}_1, \ldots,
                        \complement{\nR}_b\}$.
        For each $\complement{\nR}_i$
        already connected to some non-max-child of $z$,
        $\Gz$ has an edge between
        $z'$ and $\complement{\nR}_i$
        if $\mast(W^{z'}, \complement{R}_i) > 0$.
        Among all other $\complement{\nR}_i$,
        we pick
        $\complement{\nR}_{i'}$ and $\complement{\nR}_{i''}$
        such that $\mast(W^{z'}, \complement{\nR}_{i'})$
        and $\mast(W^{z'}, \complement{\nR}_{i''})$
        are the first and second largest.
\item Every edge $(u, \complement{\nR}_i)$ in $\Gz$ is given
        a weight of $\mast(W^u, \complement{\nR}_i)$.
\end{itemize}

\begin{lemma}
 For all $i \in [1,b]$,
        $\mwm(\Gminus{\complement{\nR}_i}) =
        \mwm(G_{z,i})$.
        Furthermore,
        $\Gz$
        can be built in
        $O((\kappa(z) + 1)\log^2 n)$ time.
\end{lemma}
\begin{proof}
The fact that
        $\mwm(\Gminus{\complement{\nR}_i}) =
        \mwm(G_{z,i})$ follows from
        the construction of $\Gz$.
Note that
$\Gz$ contains
$O(\kappa(z) + 1)$ edges.
All edges in ${\Gz}$, except
$(z',\complement{R}_{i'})$ and
$(z',\complement{R}_{i''})$,
can be found using $O(\kappa(z))$ time.
The weight of these edges can be found in $O(\kappa(z) \log n)$ time
using Lemma~\ref{EasyValues}.
To identify
$(z', \complement{R}_{i'})$ and $(z', \complement{R}_{i''})$,
        note that at most $\kappa(z)$ instances of
        $\complement{\nR}_i$ are
        connected to some non-max-child
        of $z$.  All other $\complement{\nR}_i$
        are partitioned
        into at most $\kappa(z) + 1$ intervals.
        For each interval, say $I \subseteq [1, b]$,
        by Lemma~\ref{EasyValues},
        the corresponding $\mast(W^{z'}, \complement{\nR}_i)$
        which attains the maximum in the set
        $\{ \mast(W^{z'}, \complement{\nR}_j) \mid j \in I \}$
        can be found in $O(\log^2 n)$ time.
        Thus, by scanning all the $\kappa(z)+1$ intervals,
        $\complement{\nR}_{i'}$ can be found
        in $O((\kappa(z)+1) \log^2 n)$ time.
        $\complement{\nR}_{i''}$ can be found similarly.
\end{proof}

Since $\Gz$ contains
$O(\kappa(z) +1)$ edges, and
each edge has weight  at most $N$,
we use the Gabow-Tarjan algorithm \cite{GaTa} to
compute $\mwm(\Gz)$ in $O(\sqrt{\kappa(z)+1}(\kappa(z)+1)\log N)$
time.  Then,
using our algorithm for all-cavity maximum weight matching,
we can
compute $\mwm(\Gminus{\complement{\nR}_i})$ for all $i \in [1,b]$,
and store the results
in a sparse array $A_z$
in the same amount of time.

Thus, all ${\Gz}$ with $z \in W$ can be constructed in time $\sum_{z
\in W} O((\kappa(z)+1)\log^2 n)$, which is $O(n^{1.5} \log N)$
as $\sum_{z \in W}\kappa(z) = O(n \log n)$ \cite{fpt95}. Given all
$\Gz$, the time for computing $A_z$ for all $z \in W$ is $O(\sum_{z
\in W} (\kappa(z)+1)^{1.5} \log N)$.

\begin{lemma}
$\sum_{z \in W} (\kappa(z)+1)^{1.5} \log N = O(n^{1.5} \log N)$.
\end{lemma}
\begin{proof}
Let $T(W) =
 \sum_{z \in W}
         (\kappa(z)+1)^{1.5} \log N$.
 Let $P$ be a path starting from the root of $W$
 such that every next node
 is the max-child of its predecessor.
 Then $\sum_{z \in P} \kappa(z) \leq |W| \leq n$.
 Let $\chi(P)$ denote the set
 of subtrees attached to some node on $P$.
 The subtrees in $\chi(P)$ are label-disjoint
 and each has size at most $n/2$.
Thus,
\begin{eqnarray*}
T(W) & \leq &  \sum_{z \in P} (\kappa(z)+1)^{1.5} \log N +
\sum_{W' \in \chi(P)} T(W') \\
&\leq & n^{1.5} \log N + \sum_{W' \in \chi(P)} T(W') \\
&= & O(n^{1.5} \log N).
\end{eqnarray*}
\end{proof}

\section{Auxiliary information for \mathversion{bold}$X$ with one or two
  shrunk leaves}
\label{sec_aux_1_2}

\subsection{\mathversion{bold}$X$ has  one shrunk leaf}
\label{subsec_aux_1}
Consider the computation of $\mast(W, X)$.
According to the algorithm, $\mast(W, X)$
will spawn $b$ subproblems
$\mast(W_1,X_1)$,$\ldots$, $\mast(W_b,X_b)$, which
are defined by an internal node $y$ in $X$ adjacent to the
nodes $v_1, \ldots, v_b$.
Also, for every $i \in [1,b]$, $R_i$ and $\complement{R}_i$ denote
the subtrees $X^{v_i y}$ and $X^{y v_i}$, respectively.
Suppose that $X$ has one shrunk leaf and without loss of
generality, assume that the shrunk leaf of $X$ is in
$\complement{R}_b$, i.e., $X_b$ has two shrunk leaves and all the
other $X_i$ have one shrunk leaf each.
This section shows how to find the auxiliary information required by
$W_1, \ldots, W_b$ in $O(n^{1.5} \log N)$ time.

\begin{lemma} \label{lem-one-shrunk-simple-aux}
The auxiliary information required by $W_1, \ldots, W_{b-1}$
can be computed in $O(n^{1.5} \log N)$ time.
\end{lemma}
\begin{proof}
Note that $\mast(W_1, X_1),
\ldots, \mast(W_{b-1}, X_{b-1})$ are
almost identical to the subproblems considered in
\S\ref{sec_aux_0} in that
all the $X_i$ have exactly one shrunk leaf each.
Using exactly the same approach,
we can compute the auxiliary information
in $W_1$, $\ldots$, $W_{b-1}$.
\end{proof}

The remaining section focuses on
computing the auxiliary information in $W_b$.
Let $\gamma_1$ and
$\gamma_2$ be the two shrunk leaves of $X_b$.
Assume that
$\gamma_1$ is also a shrunk leaf in
$X$,
and $\gamma_2$ represents $R_b$.
Let
$Q^+$ be the subtree obtained by
connecting $\gamma_1$ and $R_b$ together
with a node.
To compute the auxiliary information in $W_b$,
we require the values
$\mast(W^v,\gamma_1)$, $\mast(W^v,R_b)$, and $\mast(W^v, Q^+)$
for all nodes $v \in W$. 
These values are computed 
based on the following lemma.

\begin{lemma} \label{lem-values-required-one-shrunk-leaf}
$\mast(W^v,\gamma_1)$, $\mast(W^v,R_b)$, and $\mast(W^v, Q^+)$ for
all nodes $v \in W$
can be computed in $O(n^{1.5} \log N)$ time.
\end{lemma}
\begin{proof}
By Lemma~\ref{ComputeMast}, 
$\mast(W, R_b)$ and $\mast(W,Q^+)$ can be computed in time
$O(n^{1.5} \log N)$
and afterwards,
for each node $v \in W$,
$\mast(W^v, R_b)$ and
$\mast(W^v, Q^+)$
can be retrieved
in $O(1)$ time.
For each node $v \in W$, $\mast(W^v, \gamma_1)$ is
the auxiliary information stored at $v$ in $W$ and
can be retrieved in $O(1)$ time.
\end{proof}

Now, we are ready to compute the auxiliary information stored
at each node $v\in W_b$.
No auxiliary information is required for atomic leaves. Below,
Lemma~\ref{lem-internal-aux} and Lemma~\ref{lem-leaves-aux}
show that using $O(n)$ additional time, we can compute the
auxiliary information in internal nodes and
in compressed leaves, respectively.
In summary, the auxiliary information in $W_1, \ldots, W_b$
can be computed in $O(n^{1.5} \log N)$ time.

\begin{lemma} \label{lem-internal-aux}
Given $\mast(W^v,\gamma_1)$, $\mast(W^v,R_b)$, and $\mast(W^v, Q^+)$ for
all nodes $v\in W$,
the auxiliary information stored at the internal nodes in $W_b$ can be found
in $O(n)$ time.
\end{lemma}
\begin{proof}
Let $J_b$ be the set of labels of the atomic leaves of $W_b$.
An internal node $v$ can be either an auxiliary
node, a compressed node, or a node of $W| J_b$.
If $v$$ \in W|J_b$, then $v \in W$.
Thus, $\alpha_1(v) = \mast(W^v, \gamma_1)$ and
$\Wd\alpha_2(v) = \mast(W^v, R_b)$.

If $v$ is a compressed node, then
we need to compute
$\alpha_1(v), \alpha_2(v)$
and $\alpha_+(v)$.
Recall that
$v$ represents
some tree path
$\sigma=v_1,\ldots,v_k$
of $W$, where $v_1$ is the closest to the root, i.e., $v=v_1$.
Thus,
$\alpha_1(v) = \mast(W^{v_1}, \gamma_1)$,
$\alpha_2(v) = \mast(W^{v_1}, R_b)$,
and $\alpha_+(v) = \mast(W^{v_1}, Q^+)$.

Thus, $O(n)$ time is sufficient for finding
the auxiliary information stored at every internal
node of $W_b$.
\end{proof}

\begin{lemma} \label{lem-leaves-aux}
Given $\mast(W^v,\gamma_1)$, $\mast(W^v,R_b)$, and $\mast(W^v, Q^+)$ for
all nodes $v \in W$,
the auxiliary information stored at the compressed leaves in $W_b$ can be found
in $O(n)$ time.
\end{lemma}
\begin{proof}
If $v$ is a compressed leaf in $W$,
$v$'s parent $u$
must not be an auxiliary node.
Depending on whether $u$ is a compressed node,
we have two cases.

{\it Case A}:
$u$ is not a compressed node.
We must compute
$\Wd\alpha_1(v),
\Wd\alpha_2(v),
\Wd\alpha_{+}(v),
\beta(v)$.
Note that $u$ is also in $W$.
When $W_b$ is constructed from $W$,
some of the
subtrees of $W$ attached to $u$
are replaced by $v$ and no longer exist in $W_b$.
Let $W^{p_1}, \ldots, W^{p_k}$ be these subtrees.
Observe that both $v$ and $W^{p_1},
\ldots ,W^{p_k}$ represent the same set of subtrees
in $T_1$. Thus,
\begin{itemize}
\item $\Wd\alpha_1(v) =   \max\{ \mast(W^{p_i}, \gamma_1)\mid
                                1 \leq i \leq k \}$;
\item $\Wd\alpha_2(v) =
         \max\{ \mast(W^{p_i}, R_b) \mid 1 \leq i \leq k\}$;
\item $\Wd\alpha_{+}(v)  =
        \max\{ \mast(W^{p_i}, Q^+) \mid 1 \leq i \leq k\}$;
\item $\beta(v)  =  \max
        \{ \mast(W^{p_i}, \gamma_1) + \mast(W^{p_j}, R_b)
                \mid 1 \leq i \not= j \leq k \}$.
\end{itemize}
These four values can be
found in $O(k)$ time.
Since $W^{p_1}, \ldots ,W^{p_k}$ are subtrees attached to $u$ in $W$,
$k$ is at most the degree of $u$ in $W$.
Moreover, the sum of the degrees of all internal nodes
of $W$ is $O(n)$.
Therefore,
$O(n)$ time suffices to compute
the auxiliary information for all the compressed leaves
in $W_b$ whose parents are not compressed node.

{\it Case B}: $u$ is a compressed node.
We need to compute
$\alpha_1(v)$, $\alpha_2(v)$, $\alpha_+(v)$,
$\beta(v)$, $\beta_{1 \succ 2}(v)$ and
$\beta_{1 \succ 2}(v)$.
Note that $u$ is compressed from a tree path
$p_1,\ldots,p_k$
in $W$ where $p_1$ is the closest to the root.
Moreover, $v$ is compressed from the subtrees
hanging between $p_1$ and $p_k$.
For every $i \in [1,k]$, let
$\tree_i$ be the set of subtrees of $W$ attached to $p_i$
that are compressed into $v$.
Both $v$ and the subtrees in
$\cup_{1 \leq i \leq k} \tree_i$
represent the same set of subtrees in $T_1$.  The
auxiliary information stored at $v$ can be expressed as
follows.
\begin{itemize}
\item $\Wd\alpha_1(v) = \max \{
                   \mast(W^q, \gamma_1) \mid W^q \in \tree_i
                   \mbox{ for some } i \in [1,k] \}$.
\item $\Wd\alpha_2(v) = \max \{
                   \mast(W^q, R_b) \mid W^q \in \tree_i$
                    for some  $i \in [1,k] \}$.
\item $\Wd\alpha_+(v) = \max \{
                   \mast(W^q, Q^+) \mid W^q \in \tree_i $
                    for some $ i \in [1,k] \}$.
\item $\Wd\beta(v) = \max_{1 \leq i \leq k}
                     [\max \{ \mast(W^q, \gamma_1) +
                        \mast(W^{q'}, R_b) \mid  W^q, W^{q'}
                        \in \tree_i \}]$.
\item $\Wd\beta_{1 \succ 2}(v)  =
                   \max_{1 \leq j < i \leq k} [
                    \max \{ \mast(W^q, \gamma_1) \mid W^q \in
                                \tree_i \} +
                    \max \{ \mast(W^{q'}, R_b)\mid W^{q'} \in
                                \tree_j \} ]$.
\item $\Wd\beta_{2 \succ 1}(v)  =
                   \max_{1 \leq j < i \leq k} [
                    \max \{ \mast(W^{q}, R_b) \mid W^q \in\tree_i\}+
                    \max \{ \mast(W^{q'}, \gamma_1) \mid W^{q'} \in
                                \tree_j \} ]$.
\end{itemize}
These values can be found in $O(\sum_{1 \leq i \leq k} d_{p_i})$ time,
where $d_{p_i}$ is the degree of $p_i$ in $W$.  Thus, the auxiliary
information for every compressed leaf of $W_b$, whose parents are compressed
nodes, can be computed in $O(n)$ time.
\end{proof}

\subsection{\mathversion{bold}$X$ has two shrunk leaves}
\label{subsec_aux_two}
Recall that the subproblems $\mast(W_1,X_1)$, $\ldots$,
$\mast(W_b,X_b)$ are spawned from $\mast(W,X)$.
This section considers the case where $X$ has two shrunk leaves.
Without loss of generality, assume that the two shrunk leaves are in
$\complement{R}_{b-1}$ and $\complement{R}_b$, respectively.  Then,
$X_1, \ldots$, $X_{b-2}$ each have one shrunk leaf. $X_{b-1}$ and
$X_{b}$ each have two shrunk leaves. Below, we show how to
compute the auxiliary informations of $W_1, \ldots, W_b$
in $O(n^{1.5} \log N)$ time.

\begin{lemma}
The auxiliary information required by $W_1, \ldots, W_{b-2}$
can be computed in $O(n^{1.5} \log N)$ time.
\end{lemma}
\begin{proof}
The proof of this lemma is the same as that of
Lemma~\ref{lem-one-shrunk-simple-aux}.
\end{proof}

For the remaining subproblems $\mast(W_{b-1},X_{b-1})$ and $\mast(W_b,X_b)$,
both $X_{b-1}$ and $X_b$ have two shrunk leaves.
By symmetry, it suffices to discuss the
computation of $\mast(W_b,X_b)$ only.
Lemma~\ref{lem-two-shrunk-aux} shows that the auxiliary information in
$W_b$ can be computed in $O(n^{1.5} \log N)$ time.
Therefore, the auxiliary information in $W_1, \ldots, W_b$
can be computed in $O(n^{1.5} \log N)$ time.

\begin{lemma} \label{lem-two-shrunk-aux}
The auxiliary information in $W_b$ can be computed
in $O(n^{1.5} \log N)$ time.
\end{lemma}
\begin{proof}
Let $\gamma_1$ and $\gamma_2$
be the two shrunk leaves of $X_b$. Assume that $\gamma_1$ is also a
shrunk leaf in $X$ and $\gamma_2$ represents $R_b$, i.e.,
$\gamma_2$ represents the subtree $U_2^{v_b y}$ of $T_1$.  Let $Q^+$
be the subtree obtained by connecting $\gamma_1$ and $R_b$.  By the
same argument as in
Lemma~\ref{lem-internal-aux} and \ref{lem-leaves-aux},
the auxiliary information in $W_b$ can be computed
based on the values $\mast(W^v, \gamma_1)$, $\mast(W^v, R_b)$ and
$\mast(W^v,Q^+)$ for all $v \in W$. The value $\mast(W^v,\gamma_1)$
can be found in $W$.  The values $\mast(W^v, R_b)$ and
$\mast(W^v,Q^+)$ for all $v \in W$ can be retrieved in $O(1)$ time
after $\mast(W, R_b)$ and $\mast(W,Q^+)$ are computed in $O(n^{1.5}
\log N)$ time based on Lemma~\ref{lcclemma}.  Then the auxiliary
information in $W_b$ can be computed in $O(n)$ time.
\end{proof}

\section{Extension}\label{SectionMixedtree}
We have presented an $O(N^{1.5} \log N)$-time algorithm for computing
a maximum agreement subtree of two unrooted evolutionary trees of at
most $N$ nodes each.  This algorithm can be modified slightly to
compute a maximum agreement subtree for two mixed trees $M_1$ and $M_2$.

For a mixed tree $M$, a node $\ell$ is {\em consistent} with a
node $u$ if the directed edges on the path between $u$ and $\ell$ all point
away from $u$.
Let $M^u$ be the rooted tree constructed by assigning $u$ in $M$
as the root and removing the nodes of $M$
inconsistent with $u$. Given two mixed tree $M_1$ and $M_2$, we define
a maximum agreement subtree of $M_1$ and $M_2$ to be the one
with the largest number of labels among the maximum agreement subtree
of $M_1^u$ and $M_2^v$ over all nodes $u \in M_1$ and $v \in M_2$.
That is,
\[
\mast(M_1, M_2) = \max\{ \mast(M_1^u, M_2^v) \mid
        u \in M_1, v \in M_2 \}.
\]

As in the unrooted case, to compute $\mast(M_1, M_2)$,
we find a separator $y$ of $M_1$ and compute $\mast(M_1^y, M_2)$.
However, we need to delete the nodes of $M_1$ not in $M_1^y$.  When
computing $\mast(M_1^y, M_2)$, we construct some rooted subtrees of
$M_2$.  Again, we delete the nodes of $M_2$ not in these rooted
subtrees.  Such deletions are straightforward and do not increase the
time complexity of computing $\mast(M_1, M_2)$.  Thus,
$\mast(M_1,M_2)$ can be computed in $O(N^{1.5} \log N)$ time.

\section*{Acknowledgments}
The authors thank the referees for helpful comments.

\appendix

\section{Computing \mathversion{bold}\kmast($W_1,W_2$)}\label{sec_app}
\newcommand{\ordchildren}{{\rm Ch}_{\rm o}} 
Let $T_1$ and $T_2$ be rooted evolutionary trees.  Let $R_1$ and $R_2$ be
two label-disjoint rooted subtrees of $T_2$.  Let $W_1 =
\lc{T_1}{(R_1, R_2)}$ and $W_2 = \mc{T_2}{(R_1,R_2)}$.  
This section shows that $\mast(W_1,W_2)$ can be computed as if 
$W_1$ and $W_2$ were ordinary rooted
evolutionary trees \cite{fpt95,ftfocs,Kao.Evol.Siam} with some special
procedures on handling compressed and shrunk leaves.  
Note that the case where $W_1$ and $W_2$ are compressed
and shrunk with respect to a subtree can be treated as the special case
where  $R_1$ is empty.

\begin{lemma}
We can compute $\mast(W_1, W_2)$ in $O(n^{1.5} \log N)$ time,
where $n = \max \{ |W_1|, |W_2| \}$ and $N = \max \{|T_1|, |T_2|\}$.
Then, we can retrieve $\mast(W_1^u,W_2)$ for any node $u$ of
$W_1$ in $O(1)$ time.
\end{lemma}
\begin{proof}
We adopt the framework of Farach and Thorup's algorithm \cite{ftfocs},
which is essentially a sparsified dynamic programming
based on the following formula.
For any internal nodes $u$ of $W_1$ and $v$ of $W_2$,
\begin{eqnarray} \label{dynamic-prog}
\MAST(W_1^u,W_2^v) = \max
\left\{
\begin{array}{l}
\max\{ \MAST(W_1^{x}, W_2^{v}) \mid x \mbox{ is a child of }u\}; \\
\max\{ \MAST(W_1^{u}, W_2^{y}) \mid y \mbox{ is a child of }v\}; \\
\mbox{\rm r-mast}(W_1^{u}, W_2^{v}),
\end{array}
\right.
\end{eqnarray}
where $\rrmast(W_1^{u}, W_2^{v})$ denotes the maximum size of all the
agreement subtrees of $W_1^{u}$ and $W_2^{v}$ in which
$u$ is mapped to $v$.  

Our algorithm differs from Farach and Thorup's algorithm in the way
how each individual $\mast(W_1^u, W_2^v)$ is computed.
When $W_1$ and $W_2$ are ordinary evolutionary trees,
each $\mast(W_1^u, W_2^v)$ is found by computing
a maximum weight matching of some bipartite graph,
and it takes $O(n^{1.5}\log n)$ time to compute $\mast(W_1,W_2)$.  
Below, we show that when $W_1$ and $W_2$ have 
compressed and shrunk leaves, each $\mast(W_1^u, W_2^v)$ can
be found either in constant time
or by computing at most two maximum weight bipartite
matchings of similar graphs but with 
edge weights bounded by $N$ instead of $n$.  
Thus, we can compute $\mast(W_1,W_2)$
using the same sparsified dynamic programming in \cite{ftfocs};
as a by-product, we can afterwards retrieve $\mast(W_1^u, W_2)$
for any node $u$ of $W_1$ in $O(1)$ time.
The enlarged upper bound of edge weights 
increases the time complexity to $O(n^{1.5}\log N)$, though.

In the rest of this section, we show how each $\mast(W_1^u,W_2^v)$
is computed.
First, we consider the case when $u$ is a leaf. The following
case analysis shows that $O(1)$ time
suffices to compute $\mast(W_1^u,W_2^v)$.

{\it Case 1}: $u$ is an atomic leaf. 
If $W_2^v$ contains a leaf with the same label as that of $u$, then
$\mast(W_1^u, W_2^v)=1$; otherwise it equals zero.

{\it Case 2}: $u$ is an auxiliary leaf. Then, $\mast(W_1^u, W_2^v)=0$.

{\it Case 3}: $u$ is a compressed leaf.
By definition, $u$ can only be mapped to $\cleaf_1$, $\cleaf_2$ or
the least common ancestor $y_c$ of $\cleaf_1$ and $\cleaf_2$.
If $W_2^v$ has no shrunk leaves, then $\mast(W_1^u, W_2^v) =0$.
If $W_2^v$ has only one shrunk leaf, say $\cleaf_1$, then
$\mast(W_1^u, W_2^v)=\ralpha{1}{u}$. If $W_2^v$ has
two shrunk leaves, $W_2^v$ must also contain $y_c$ and
$\mast(W_1^u, W_2^v) = \max\{ \ralpha{1}{u},
\ralpha{2}{u}, \ralpha{+}{u} \}$.

Next, we consider the case when $u$ is an internal node.
Assume that $v$ is an atomic leaf. Then $\mast(W_1^u, W_2^v)=1$
if $W_1^u$ contains a leaf with the same label as that of $v$,
and zero otherwise.
If $v$ is a shrunk leaf,
say $\cleaf_1$, then $\mast(W_1^u, W_2^v) = \ralpha{1}{u}$.
It remains to consider the case when $v$ is an  internal node.
Due to the nature of dynamic programming, 
we only need to compute $\rrmast(W_1^u, W_2^v)$, then
we can apply the Equation (\ref{dynamic-prog}) to compute $\mast(W_1^u,W_2^v)$.
We further divide our discussion into the following three cases.

{\em Case 1}: $u$ is an auxiliary internal node.
In such case, $u$ has only two children, one of them is
an auxiliary leaf.  From definition, an auxiliary
leaf will not be mapped to any node in any agreement subtree of
$W_1^u$ and $W_2^v$; thus, there is no agreement
subtree in which $u$ is mapped to $v$
and $\rrmast(W_1^u, W_2^v)=0$.

{\em Case 2}: $u$ is an ordinary internal node.
As in \cite{ftfocs}, we first construct
the bipartite graph defined as follows:
Let $A$ and $B$ be the set of children of $u$ 
and $v$, respectively;
define $G[A,B]$ to be
the bipartite graph formed by the edges $(x,y) \in A \times B$ with
$\mast(W_1^x, W_2^y) > 0$, and $(x,y)$ is given a weight
$\mast(W_1^x, W_2^y)$. 

If none of $u$'s children is an auxiliary leaf, 
then $\rrmast(W_1^u, W_2^v)=\mwm(G[A,B])$.
Otherwise, let $\bar{z}$ be the child of $u$ which
is an auxiliary.  In this case, $u$ also has a compressed
child $z$.  Other than $z$ and $\bar{z}$, no other child of $u$
is a compressed and auxiliary leaf.
On the other hand, consider the rooted subtrees $W_2^x$
rooted at the children $x$ of $v$.
If the shrunk leaves appear together in one of such subtrees,
then by definition,  $\bar{z}$ cannot be mapped to any shrunk leaf
in any agreement subtree of $W_1^u$ and $W_2^v$, and 
$\rrmast(W_1^u, W_2^v)=\mwm(G[A,B]-\{\bar{z}\})$.  
If the shrunk leaves appear in two different subtrees
rooted at two children  $y_1$ and $y_2$ of $v$, then
\[ \rrmast(W_1^u, W_2^v) = \max \Big \{
      \mwm(G[A,B] - \{\bar{z}\}), \mwm(G[A, B]-\{z,\bar{z}, y_1,y_2\}) + 
       \beta(z)\Big \}.
\]

{\em Case 3}: $u$ is a compressed internal node.
By definition of a compressed node, the structure of $W_1^u$ is very
restrictive---$u$ has exactly three children
$z$, $\bar{z}$, and an auxiliary internal node $\bar{u}$; $\bar{u}$
has two children, an auxiliary leaf $\bar{\bar{z}}$ and an
uncompressed internal node $w$; see Figure~\ref{structure}.  To find
\begin{figure}
\begin{center}
\begin{picture}(0,0)%
\epsfig{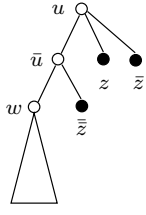}%
\end{picture}%
\setlength{\unitlength}{0.0008400in}%
\begingroup\makeatletter\ifx\SetFigFont\undefined
\def\x#1#2#3#4#5#6#7\relax{\def\x{#1#2#3#4#5#6}}%
\expandafter\x\fmtname xxxxxx\relax \def\y{splain}%
\ifx\x\y   
\gdef\SetFigFont#1#2#3{%
  \ifnum #1<17\tiny\else \ifnum #1<20\small\else
  \ifnum #1<24\normalsize\else \ifnum #1<29\large\else
  \ifnum #1<34\Large\else \ifnum #1<41\LARGE\else
     \huge\fi\fi\fi\fi\fi\fi
  \csname #3\endcsname}%
\else
\gdef\SetFigFont#1#2#3{\begingroup
  \count@#1\relax \ifnum 25<\count@\count@25\fi
  \def\x{\endgroup\@setsize\SetFigFont{#2pt}}%
  \expandafter\x
    \csname \romannumeral\the\count@ pt\expandafter\endcsname
    \csname @\romannumeral\the\count@ pt\endcsname
  \csname #3\endcsname}%
\fi
\fi\endgroup
\begin{picture}(1113,1284)(303,-738)
\put(971, -8){\makebox(0,0)[b]{\smash{\SetFigFont{8}{9.6}{rm}$z$}}}
\put(1186, -8){\makebox(0,0)[b]{\smash{\SetFigFont{8}{9.6}{rm}$\bar{z}$}}}
\put(831,-293){\makebox(0,0)[b]{\smash{\SetFigFont{8}{9.6}{rm}$\bar{\bar{z}}$}}}
\put(411,-163){\makebox(0,0)[b]{\smash{\SetFigFont{8}{9.6}{rm}$w$}}}
\put(686,462){\makebox(0,0)[b]{\smash{\SetFigFont{8}{9.6}{rm}$u$}}}
\put(551,137){\makebox(0,0)[b]{\smash{\SetFigFont{8}{9.6}{rm}$\bar{u}$}}}
\end{picture}
\end{center}
\caption{Structure of $W_1^u$.}
\label{structure}
\end{figure}
$\rrmast(W_1^u, W_2^{v})$, we note that there are only
six possible ways on how
the $z, \bar{z}, \bar{\bar{z}}$ are mapped to 
$\cleaf_1$ and $\cleaf_2$.  We consider each of these
cases and $\rrmast(W_1^u,W_2^v)$ is the maximum of the 
values found.
We only discuss the case where $\cleaf_1$ and $\cleaf_2$
are mapped to
$z$ and $\bar{\bar{z}}$, respectively.
The other cases can be handled similarly.
Let ${P}$ be the path between $\gamma_2$ and 
$y_c$.  Let $\Stree({P})$ denote the set of
subtrees hanged on ${P}$.  The size of the largest
agreement subtrees of $W_1^u$ and $W_2^v$ in which $\cleaf_1$ and
$\cleaf_2$ are
mapped to $z$ and $\bar{z}$, respectively, equals
\begin{equation}\label{eqn}
 \beta_{1
\succ 2}(z) + \max \{ \mast(W_1^w, \tau) \mid \tau \in \Stree({P})
\}.
\end{equation}
Note that using the technique in \cite{ftfocs}, we can precompute
$\max \{ \mast(W_1^{x}, \tau) \mid \tau\in \Stree({P}) \}$ for
all $x \in W_1$ in $O(n^{1.5} \log N)$ time. Afterwards,
(\ref{eqn}) can be found in constant time.
\end{proof}

\newcommand{\pivotseq}{\Gamma}
\newcommand{\maxrange}{\alpha}
\newcommand{\maxran}{\beta}
\newcommand{\descendent}{\beta}

\section{Preprocessing for finding
        \mathversion{bold}$\max\{A_z[i] | z \in W^v\}$}\label{app_b}

Let $h$ be the number of nodes in $W$.
Consider the $h$ arrays $A_z$ of dimension $b$ where $z \in W$.
Recall that $m_z$ is the number of distinct values in $A_z$, and $m =
\sum_{z \in W} (m_z + 1)$.
This section describes an $O(m \alpha(h))$-time preprocessing, which
supports finding $\max \{A_z[i] \mid z \in W^v \}$, for any $i \in
[1,b]$ and any node $v$ of $W$, in $O(\log h)$ time.

By definition, each $A_z$ has at least $b-m_z$ entries storing some
common value $c_z$.  For every $i \in [1,b]$, let $\pivotseq_i$ be the
set of nodes $z$ where $A_z[i]$ stores a value different from $c_z$.
Note that $\sum m_z = \sum_{1 \leq i \leq b} |\pivotseq_i|$.  We
assume that each node of $W$ is identified uniquely by an integer in
$[1,h]$ assigned by a preorder tree traversal \cite{clr}.  For any
node $v \in W$, let $\descendent(v)$ be the number of proper
descendents of $v$ in $W$.

Based on Lemma~\ref{equal}, $\max \{A_z[i] \mid z \in W^v\} =
\max \{ A_z[i] \mid z \in [v, v+\beta(v)] \}$ for any $v \in W$,
$i \in [1, b]$.
Therefore, to solve our problem, it is sufficient to give
an $O(m\alpha(h))$-time preprocessing to support finding
$ \max\{ A_z[i] \mid z \in H \}$ for any $i \in [1,b]$ and any
interval $H \subseteq [1,h]$ in $O(\log h)$ time.
\begin{lemma} \label{equal}
For any $v \in W$ and $i \in [1,b]$,
$
        \max \{ A_z[i] \mid z \in W^v \} =
        \max \{ A_z[i] \mid z \in  [v, v+\descendent(v)] \}.
$
\end{lemma}
\begin{proof}
Straightforward.
\end{proof}

Our preprocessing does not work on each sequence
$A_1[i],A_2[i],\ldots,A_h[i]$ directly.  Instead, it first draws
out useful information about the common values $c_z$ stored in the
sequences and applies a contraction technique to shorten each
sequence.  Then, it executes Fact~\ref{fact_noba} on these shortened
sequences.

\begin{fact}[see \cite{Noba}]\label{fact_noba}
Given any sequence $a_1, \ldots,
a_h$ of real numbers, we can preprocess
these $h$ numbers in
$O(h)$ time so that we can find
the maximum of any subsequence
$a_x,a_{x+1}, \ldots, a_y$ in $O(\alpha(h))$ time.
\end{fact}

Our preprocessing is detailed as follows. Its time complexity
is $O(m \alpha(h))$ as shown in Lemma~\ref{lem-preprocess-time}.

\begin{enumerate}
\item
        For each $i \in [1,b]$,
        find $\pivotseq_i$ and
        arrange the integers in $\pivotseq_i$
        in ascending order.
\item
        Apply Fact~\ref{fact_noba}
        to the sequence $c_1, \ldots, c_{h}$.
\item
        For every non-empty $\pivotseq_i =
        \{ x_1 < \ldots < x_d \}$,
        compute
        $\maxran_\ell =
                \max \{ A_x[i] \mid x \in (x_{\ell}, x_{\ell+1})\}$
        for every $\ell \in [1,d-1]$,
                and then
        apply Fact~\ref{fact_noba}
        to the sequence
        $A_{x_1}[i], \maxran_1, A_{x_2}[i], \ldots,
                \maxran_{d-1}, A_{x_d}[i].$
\end{enumerate}
\begin{lemma} \label{lem-preprocess-time}
The preprocessing requires $O(m \alpha(h))$ time.
\end{lemma}
\begin{proof}
We can examine all the entries of $A_z$ whose values differ from
$c_z$ in $O(m_z)$ time.  By examining all such entries of
$A_1,\ldots,A_h$, we can construct $\pivotseq_i$ and arrange the
integers in $\pivotseq_i$ in ascending order.  Thus, Step~1 takes
$O(m)$ time.  Step~2 takes $O(h)=O(m)$ time.  For Step 3, we first
analyze the time required to process one nonempty $\pivotseq_i=\{x_1 <
\ldots < x_d\}$.  Note that $(x_{\ell}, x_{\ell+1}) \cap \pivotseq_i =
\phi$ for every $\ell \in [1,d-1]$. Thus, $\maxran_\ell = \max\{c_x
\mid x \in (x_{\ell}, x_{\ell+1})\}$ can be computed in
$O(\alpha(n))$ time using the result of Step 2.  Summing over all
$\ell \in [1, d-1]$, computing all $\maxran_\ell$ takes
$O(|\pivotseq_i|\alpha(h))$ time.  Applying Fact~\ref{fact_noba} to
the sequence $A_{x_1}[i]$, $\maxran_1$, $A_{x_2}[i]$, $\ldots$
$\maxran_{d-1}$, $A_{x_d}[i]$ takes $O(|\pivotseq_i|)$ time. In total,
it takes $O(|\pivotseq_i|\alpha(h))$ time to process one
$\pivotseq_i$, and Step 3 takes $O(\sum |\pivotseq_i|\alpha(h))
=O(m\alpha(h))$ time.  Thus, the total time of our preprocessing is
$O(m \alpha(h))$.
\end{proof}

After the preprocessing, each query can be answered in $O(\log h)$ time
as stated in the following lemma.
\begin{lemma}
After the preprocessing, $\max \{ A_z[i] \mid z \in H \}$ can
be found in $O(\log n)$ time for any $i \in [1, b]$ and
any interval $H \subseteq [1, h]$.
\end{lemma}
\begin{proof}
Let $H = [p,q]$.
A crucial step is to find
$[p,q] \cap \pivotseq_i$.
Without loss of generality,
assume $\pivotseq_i \neq \phi$.
To find $[p,q] \cap \pivotseq_i$,
we first find
the smallest integer $x_s$ in $\pivotseq_i$ that
is greater than  $p$,  and the largest integer
$x_t$ in $\pivotseq_i$ that is smaller than $q$.
Since $\pivotseq_i$ is sorted,
we can
find $x_s$ and $x_t$
in $O(\log |\pivotseq_i|)= O(\log h)$ time.
If $x_s > x_t$, then
$[p,q] \cap \pivotseq_i = \phi$;  otherwise,
$[p,q] \cap \pivotseq_i$ is the set of integers
between $x_s$ and $x_t$ in $\pivotseq_j$.

If $[p,q] \cap \pivotseq_i = \phi$, then
$\max \{ A_x[i] \mid x \in [p,q] \}$ $=$
$\max \{ c_x \mid x \in [p,q] \}$.
Because of Step 1 of our preprocessing,
we can find $\max \{ c_x \mid x \in [p,q] \}$
in $O(\alpha(h))$ time.

If $[p,q] \cap \pivotseq_i = \{ x_s< x_{s+1}< \ldots< x_{t}\}$,
then
$[p,q] = [p,x_s -1] \cup \{x_s\} \cup (x_{s}, x_{s+1})$
$\cup \cdots \cup
\{x_t\} \cup [x_{t} + 1,q]$ and
$\max \{A_z[i] \mid z \in [p,q]\}$ equals the maximum of
\begin{enumerate}
\item $\max\{ A_x[i] \mid x \in [p,x_s -1]\}$,
\item $\max\{ A_x[i] \mid x \in
        \{x_s\} \cup (x_s, x_{s+1}) \cup \cdots \cup
        (x_{t-1}, x_t) \cup \{x_t \} \}$,
\item $\max\{ A_x[i] \mid x \in [x_t +1, q] \}$.
\end{enumerate}
Note that Item 2 equals the maximum of
$A_{x_s}[i], \maxran_{s},\ldots, \maxran_{t-1},
A_{x_t}[i]$, which can be computed in $O(\alpha(h))$ time
after Step 3 of our preprocessing.
Since $\pivotseq_i \cap [p,x_s -1] = \phi$ and
$\pivotseq_i \cap [x_t + 1, q] = \phi$,
Step 2 enables us to compute Items 1 and 3
in $O(\alpha(h))$ time.
As a result, $\max \{ A_z[i] \mid z \in [p,q]\}$ can be answered in
$O(\log h)$ time.
\end{proof}

\bibliographystyle{siam}
\bibliography{all}

\end{document}